\documentclass[a4paper,fleqn]{cas-dc}

\usepackage[numbers]{natbib}
\usepackage{gensymb}
\usepackage{graphicx,caption}
\usepackage{amsmath}
\usepackage{amsfonts}
\usepackage{amssymb}
\usepackage{comment}
\def\tsc#1{\csdef{#1}{\textsc{\lowercase{#1}}\xspace}}
\tsc{WGM}
\tsc{QE}
\tsc{EP}
\tsc{PMS}
\tsc{BEC}
\tsc{DE}
\begin{document}
\setlength{\mathindent}{0pt}
\let\WriteBookmarks\relax
\def\floatpagepagefraction{1}
\def\textpagefraction{.001}
\shorttitle{Robust MPC with data-driven demand forecasting for frequency regulation with heat pumps}
\shortauthors{F Bünning et~al.}

\title [mode = title]{Robust MPC with data-driven demand forecasting for frequency regulation with heat pumps}                      

\author[1,2]{Felix Bünning}
\cormark[1]
\credit{Conceptualization, Methodology, Software, Validation, Investigation, Visualization, Writing - Original Draft}

\author[2]{Joseph Warrington}
\credit{Methodology, Supervision, Writing - Review \& Editing}

\author[1]{Philipp Heer}
\credit{Supervision, Funding acquisition, Writing - Review \& Editing, Conceptualization}

\author[2]{Roy S. Smith}
\credit{Supervision, Funding acquisition, Writing - Review \& Editing, Conceptualization}

\author[2]{John Lygeros}
\credit{Supervision, Funding acquisition, Writing - Review \& Editing, Conceptualization}

\address[1]{Empa, Urban Energy Systems Laboratory, Überlandstrasse 129, 8600 Dübendorf, Switzerland}
\address[2]{Automatic Control Laboratory, Department of Electrical Engineering and Information Technology, ETH Zürich, Switzerland}

\cortext[cor1]{Corresponding author}

\begin{abstract}
With the increased amount of volatile renewable energy sources connected to the electricity grid, and the phase-out of fossil fuel based power plants, there is an increased need for frequency regulation. On the demand side, frequency regulation services can be offered by buildings or districts that are equipped with electric heating or cooling systems, by exploiting their thermal inertia. Existing approaches for tapping into this potential typically rely on dynamic building models, which in practice can be challenging to obtain and maintain. As a result, practical implementations of such systems are scarce. Moreover, actively controlling buildings requires extensive control infrastructure and may cause privacy concerns in district energy systems. Motivated by this, we exploit the thermal inertia of buffer storage for reserves, reducing the building models to demand forecasts here. By combining a control scheme based on Robust Model Predictive Control, with affine policies, and heating demand forecasting based on Artificial Neural Networks with online correction methods, we offer frequency regulation reserves and maintain user comfort with a system comprising a heat pump and buffer storage. While the robust approach ensures occupant comfort, the use of affine policies reduces the effect of disturbance uncertainty on the system state. In a first-of-its-kind experiment with a real district-like building energy system, we demonstrate that the scheme is able to offer reserves in a variety of conditions and track a regulation signal while meeting the heating demand of the connected buildings. 13.4\% of the consumed electricity is flexible. In additional numerical studies, we demonstrate that using affine policies significantly decreases the cost function and increases the amount of offered reserves and we investigate the suboptimality in comparison to an omniscient control system.

\end{abstract}

\begin{keywords}
Frequency regulation \sep Robust Model Predictive Control \sep Demand forecasting \sep Electric reserves \sep District heating \sep Affine policies 
\end{keywords}
\maketitle

\section{Introduction}
\label{sec: Introduction}

The amount of renewable energy sources in the electricity grid is continuously increasing. As many of these sources are highly volatile, there is a growing need for frequency regulation \cite{Johnson2019}. A common strategy for frequency regulation is the deployment of fast-reacting power plants, for example gas or hydro-power; an emerging strategy is the use of storage technologies, for example batteries. Besides such regulation on the supply side of the grid, frequency regulation on the demand side is possible through manipulation of controllable loads. This concept falls under the category of demand-side management \citep{Gelazanskas2014}.

Possible candidates for demand-side management are buildings equipped with electric heating or cooling systems, such as heating, ventilation and air conditioning (HVAC) units, electric heaters and heat pumps \cite{Hao2015}, or entire district heating systems with electric heat sources. Due to the thermal inertia of buildings, they are to an extent flexible when it comes to their heating and cooling requirements, hence their electricity consumption. By shifting their consumption in time they can therefore influence the grid frequency \cite{Fischer2017}.

However, shifting electricity consumption can have an impact on occupant comfort as heating and cooling energy might not be available at the exact time when it is needed. There are different strategies to mitigate this impact. The authors of \cite{Rominger2018,RomeroRodriguez2019,Maasoumy2013} develop and test control strategies for frequency regulation with heat pumps and HVAC units without explicitly enforcing comfort constraints and check only a-posteriori whether these were violated or not. References \cite{Zhao2013,Wang2020} use heuristics based on weather forecasts and occupancy to limit the offered frequency reserve capacity to enforce comfort constraints. Other authors use dynamic building models to exactly determine the influence of changed heating and cooling supply on room temperatures when providing reserves with heat pumps \citep{Kim2015} or air handling units \citep{Hao2013,Lin2013,Olama2018}. 

Combined with optimization in the frame of Model Predictive Control (MPC), such dynamic models can be used to maximize the offered frequency reserves while maintaining comfort constraints \citep{Rastegarpour2020,Rastegarpour2020a}. References \citep{Zhang2017} and \citep{Vrettos2016} develop Robust MPC schemes to provide day-ahead reserves for frequency regulation with commercial buildings and HVAC systems. Robust schemes ensure occupant comfort in the face of uncertainty in the regulation signal from the transmission system operator (TSO). In \citep{Vrettos2018,Vrettos2018a} this approach is further developed and tested in a case study on a real small air-conditioned building. In \cite{Oldewurtel2013}, MPC is used to provide ancillary services with a real building and simulation studies are conducted to investigate a setup of an aggregation of office buildings. 

As buildings and related components (such as buffer storage) have integrator dynamics, uncertain disturbances lead to uncertainty accumulation in the system state and thus limit the electrical reserves offered. This is the case because the Robust MPC uses much of the available flexibility by accounting for the uncertainty instead of exploiting it for reserves. Feedback policies are a known tool in Stochastic and Robust MPC to mitigate this problem \citep{Goulart2006}, as they allow the controller to plan to react on uncertainties before they are revealed. However, to the best of the authors' knowledge, they have not yet been applied in practice to reserve provision with buildings and districts.

The need for detailed models of the thermal dynamics of buildings is a challenge for MPC based approaches that exploit the thermal inertia of buildings. Some authors argue that the cost of developing and maintaining first-principles building models could be holding back the wide-spread application of MPC in buildings in general \cite{Sturzenegger2016,Jain2018}, which would also have implications on the use of MPC for building demand response as an extension to general MPC for building control \cite{Vrettos2018}. In addition to the grey-box modeling approach \citep{Dimitriou2015,DeConinck2015,Li2021}, where parameters of a physical model are fitted to measurement data, there is growing interest in data-driven and machine learning based building models, both of which can be more cost-effective compared to first-principles models, as they are parametrized with measurement data. However, data-driven model frameworks are so far only applicable to small and simple buildings \cite{Bunning2020,Jain2018}. Moreover, if buildings are to be controlled from an external entity in the context of reserve provision, for example by an aggregator \cite{Oldewurtel2013}, occupants might have privacy concerns related to room temperature and other live measurements, and could therefore prefer to not actively participate. Furthermore, most of today's buildings are not yet equipped with the necessary hardware for predictive room temperature control. To also account for these cases, alternative strategies for reserve provision schemes that do not necessarily rely on exploiting the thermal inertia of buildings are necessary. These strategies should, however, be extendable to exploiting building thermal dynamics for applications where the framework conditions allow it.

Here, we take a step in this direction by decoupling heating/cooling demand and heating/cooling supply of the system. This is achieved by placing a buffer storage, e.g. a water tank, between supply and demand. In this case, the thermal inertia of the buffer storage allows flexibility in heating and cooling energy production without the necessity to exploit the inertia of the connected building itself. The comfort of the occupants is ensured by a separate (potentially unknown) lower level controller as long as the buffer storage has a sufficiently high temperature. The heating/cooling system and storage can in this case be modeled with first principles, which is tractable from an economic point of view as these are mass-produced products. The demand of the building can be modeled with a forecasting method.

In contrast to dynamic building models, data-driven heating and cooling demand forecasting for buildings or whole neighbourhoods and districts is a mature field, see for example the reviews \citep{MatDaut2017,Zhao2012,Foucquier2013,Suganthi2012} for comparisons of data-driven with conventional methods, \citep{Wang2017, Amasyali2018,Ahmad2018a} for different data-driven methods, and \citep{Harish2016} with a focus on forecasts for control strategies.
Here, inputs such as weather forecasts and calendar features, for example hour of the day or workday/non-workday, are used in combination with machine learning methods, often Artificial Neural Networks (ANN), to forecast heating or cooling energy consumption on different timescales. While some authors use different forms of single ANN, for example Feed-forward Neural Networks \citep{Park2010,Saloux2018}, Recurrent Neural Networks \citep{Kato2008}, and Deep Neural Networks \citep{Suryanarayana2018} for the prediction of heating and cooling demands of districts, \cite{Paudel2014,Kwok2011,Leung2012,Mestekemper2013} do so for individual buildings. To mitigate the problem of high prediction variance between individual networks \cite{Jovanovic2015a,Jetcheva2014} and \cite{DeFelice2011} propose ensemble methods in the context of building demand prediction. As ensemble methods have the disadvantage of being computationally expensive, we have developed and validated correction methods based on online learning and error autocorrelation correction methods, which both decrease variance and increase accuracy, while avoiding the disadvantages of ensemble methods \cite{BunningFelixBollingerAndrewHeerPhilippSmithRoy2018}.

The above named configuration of a central heating/cooling device connected to buffer storage is not only common for many commercial buildings, but also for entire district heating systems \citep{Hennessy2019}. Here, central heat pumps and buffer storages are controllable by the system operator, while connected buildings might not be controllable (e.g. due to privacy concerns or hardware/model limitations). These can be forecast, but not directly influenced. Using the thermal buffer presents an option to offer electricity reserves even if the building's thermal mass cannot be controlled, due to one or more of the reasons given above. 

There is a growing interest in the potential \cite{Lund2018,Ivanova2019} of providing ancillary services in district heating systems, as witnessed by attempts to quantify flexibility potential \citep{Xu2020} and model individual components for this purpose \citep{PaghNielsen2021}. However, review studies \cite{Vandermeulen2018a,Hennessy2019} indicate that there is little work on the operation \cite{Terreros2020,Salpakari2016,Li2016} of such systems with reserve schemes and no implementation and validation of day-ahead reserves on a real district heating system in the literature. One of the main contributions of this work is therefore the demonstration of a viable approach on a real system. We feel that enabling deployment at the district level is an important step because, with the current structure of ancillary markets, individual buildings can only enter through massive aggregation \citep{Geidl2017,Koch2011,Borsche2014,Oldewurtel2013}; districts, however, may be able to enter individually, or in small clusters, if minimal power requirements are reached. Moreover, deploying the methods in districts instead of individual buildings can improve performance because of increased demand predictability thanks to averaging \citep{BunningFelixBollingerAndrewHeerPhilippSmithRoy2018} and because of synergistic effects, for example between residential and commercial buildings \citep{Darivianakis2017}.

However, for a practical application of such a reserve scheme, the combination of advanced methods is essential. High-accuracy demand forecasting methods are needed to reduce the uncertainty in the disturbance of the buffer storage, and as a result the uncertainty in the system state. Robust MPC \citep{Vrettos2016} is needed to ensure occupant comfort in the face of uncertainty of the demand forecasts, the frequency regulation signal and other disturbances. Moreover, to maximize the reserves offered, feedback policies \citep{Goulart2006} are required, to minimize the effect of disturbance uncertainty on the system state, and as a result to better exploit the available storage compared to open-loop MPC. This is especially important in the considered configuration, as the thermal inertia of buffer storage is considerably smaller compared to the inertia of the thermal mass of buildings. 

%The problem of modeling the dynamics of buildings is reinforced by privacy concerns regarding indoor temperature measurements in this setting. There is a growing interest in the potential of providing ancillary services in district heating systems \cite{Lund2018,Xu2020,PaghNielsen2021,Ivanova2019}. However, review studies \cite{Vandermeulen2018a,Hennessy2019} indicate that there is little work on the operation \cite{Terreros2020,Salpakari2016,Li2016} of such systems and no implementation and validation on real a real system in the literature.

%In conclusion, we see that the concept of using buffer storage for reserves provision in buildings or district heating systems, where supply and demand are decoupled through said buffer storage, has not been explored yet, although it has several advantages: The heating demand of connected buildings can be modeled with forecasting models instead of dynamic building models, which are easier to obtain; Moreover, information about the temperature inside the connected buildings is not necessary, which reduces privacy concerns. To the best of our knowledge, there is no demonstration of a working control concept for this decoupled system configuration in practice.

In this work we therefore combine the 3-level Robust MPC for frequency regulation approach presented in \cite{Bunning2020a} with the forecasting methods presented in \cite{BunningFelixBollingerAndrewHeerPhilippSmithRoy2018} to offer day-ahead frequency regulation reserves with a system comprising a ground-source heat pump and water buffer storage that meet the heating demand of a building or group of buildings. The robust formulation is combined with affine policies, as discussed in \cite{Warrington2012} for reserve provision in power systems, which allows us to increase the reserves offered compared to standard open-loop MPC. While the individual methods have been presented in previous work, their combination is required to achieve a practical implementation to enable the deployment of ancillary services to the level of districts with central heat supply.
We also document what we believe to be the first implementation demonstrating this approach in experiments: We validate the methods on a physical system, the NEST demonstrator in Switzerland, a ``vertical neighbourhood'' connected by a district heating system \cite{Richner2018}. We show that the approach is able to offer a substantial amount of regulation reserves in a variety of ambient conditions and on a configuration, where the storage capacity is only a small fraction of the daily heating demand. We also demonstrate that the approach ensures good regulation signal tracking performance with a variable speed heat pump. Furthermore, we investigate optimality properties of the MPC solutions and investigate the effect of the use of affine policies in two numerical experiments. The presented scheme can be extended to include building thermal dynamics in future work.

The remainder of the article is structured as follows. In Section \ref{sec: Problem statement} we introduce the reserve provision scheme and the system under consideration. In Section \ref{sec: Methodology} we discuss the models for heat pump and storage as well as the prediction models with correction methods for the heating demand of buildings. We also describe the Robust MPC based control scheme.  In Section \ref{sec: Experimental case study and results} we present the experimental case study and its results. In Section \ref{sec: Numerical case study and results} we describe the numerical case studies and discuss the suboptimality of the presented approach. We conclude in Section \ref{sec: Conclusion}.

\section{Problem statement}
\label{sec: Problem statement}

\subsection{Reserve provision scheme}
\label{subsec: Reserve provision scheme}

We assume a day-ahead planning frequency regulation reserve scheme by the regulation product RegD offered by the U.S. transmission system operator PJM. We use the PJM reserve market to have a concrete example. However, the method is not limited to this scheme and can be easily adapted to other ancillary market conditions. In the considered scheme, the reserve provider communicates an offer $r \in \mathbb{R}^{96}$ of symmetric reserves to the TSO at midnight. The offer is made in 15-minute intervals and is fixed for the next 24 hours. During the next day, when the offered reserves are due, the reserve provider can change their base consumption $u^0_{k}$ every timestep $k$ (i.e. every 15 minutes). It should then track the electrical load

\begin{equation}
u_{k}(\tau)=u^0_{k}+w(\tau) r_k,
\label{eq: reserve}
\end{equation}

\noindent where $w(\tau) \in [-1,1]$ denotes the regulation signal which is updated every 2 seconds by the TSO, and $r_k$ denotes the k$^\text{th}$ element of the list of offered reserves $r$. $u^0_{k}$ is updated every 15 minutes, $r_k$ is time varying in 15 minute intervals but fixed for the day, while $w(\tau)$ changes every 2 seconds. Thus, $u_{k}(\tau)$ also changes and is sampled every 2 seconds.

Note that the offered reserves are effectively symmetric because $w(\tau)$ varies between -1 and 1 in the considered scheme. As discussed in \citep{Vrettos2016thesis}, other schemes exist where, assymetric reserves or only up- or down-reserves are offered. The presented methods readily extend to these schemes, for example by constraining the values of $w(\tau)$ to $[-1,0]$ for only up-reserves, or by introducing separate variables $r_{k,up}$ and $r_{k,down}$ for up- and down-reserves.

The tracking performance is judged by a \textit{composite performance score} monitored by the TSO, which consists of an \textit{accuracy score}, which measures the correlation between the reserve signal and the system response, a \textit{delay score} which measures the time delay between reserve signal and system response, and a \textit{precision score} which measures the error between reserve signal and system response \cite{PJM}.

Our scheme does not consider time-varying energy prices and peak pricing for $u^0_{k}$, as these aspects have been treated in the literature already, and their handling would dilute the scope of this work. We therefore refer the interested reader to \citep{Zhang2015,Sundstrom2017,Ma2014a}.

\subsection{System under consideration}

\begin{figure}
%	\capstart
	\centering
		\includegraphics[width=0.45\textwidth]{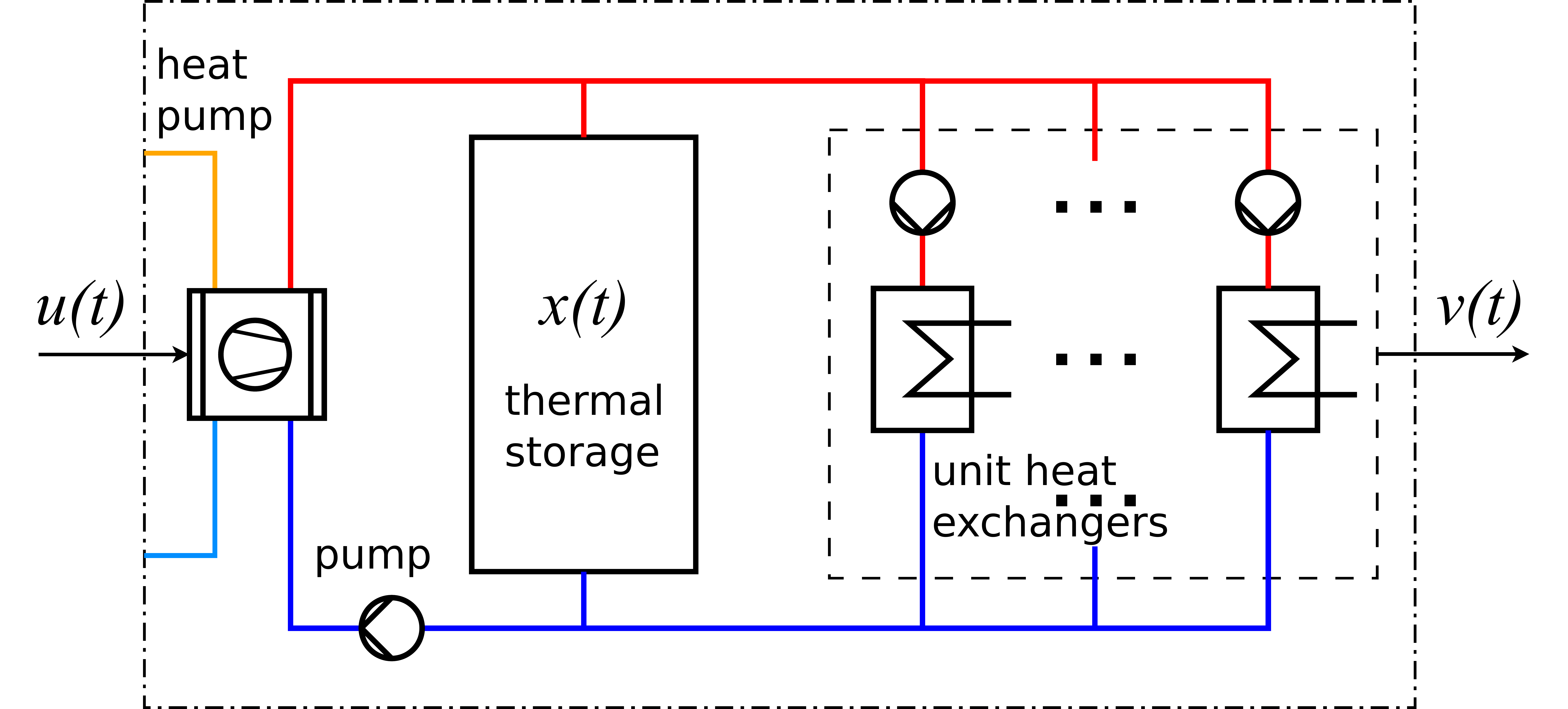}
	\caption{Schematic of the system under consideration with heat pump, water storage tank and heat exchangers for individual apartments}
	\label{fig: Schematic of the heating system with heat pump, water storage tank and heat exchanger}
\end{figure}

We consider the heating system for reserve provision shown in Figure \ref{fig: Schematic of the heating system with heat pump, water storage tank and heat exchanger}. It consists of a vapour compression cycle  heat pump, which is depicted on the left, and a water storage tank, which is depicted in the middle. The heat pump draws cold water from the bottom of the storage with the help of a pump, warms up the water by transferring heat from the refrigerant to the water inside the condenser, and feeds it back into the top of the storage. By varying the heat pump's electrical consumption, frequency regulation can be offered. On the right, individual pumps draw warm water from the top of the storage tank and pass it through heat exchangers, which supply individual apartments of a building or individual buildings of a district with heat. The cold water is returned to the bottom of the tank. Note that building-sized heat pumps would usually not be eligible for reserve provision programs as presented above due to minimum capacity requirements. Multiple buildings would therefore have to be aggregated under these programs.

\subsection{Control scheme}
\label{subsec: Control scheme}

\begin{figure}
%	\capstart
	\centering
		\includegraphics[width=0.45\textwidth]{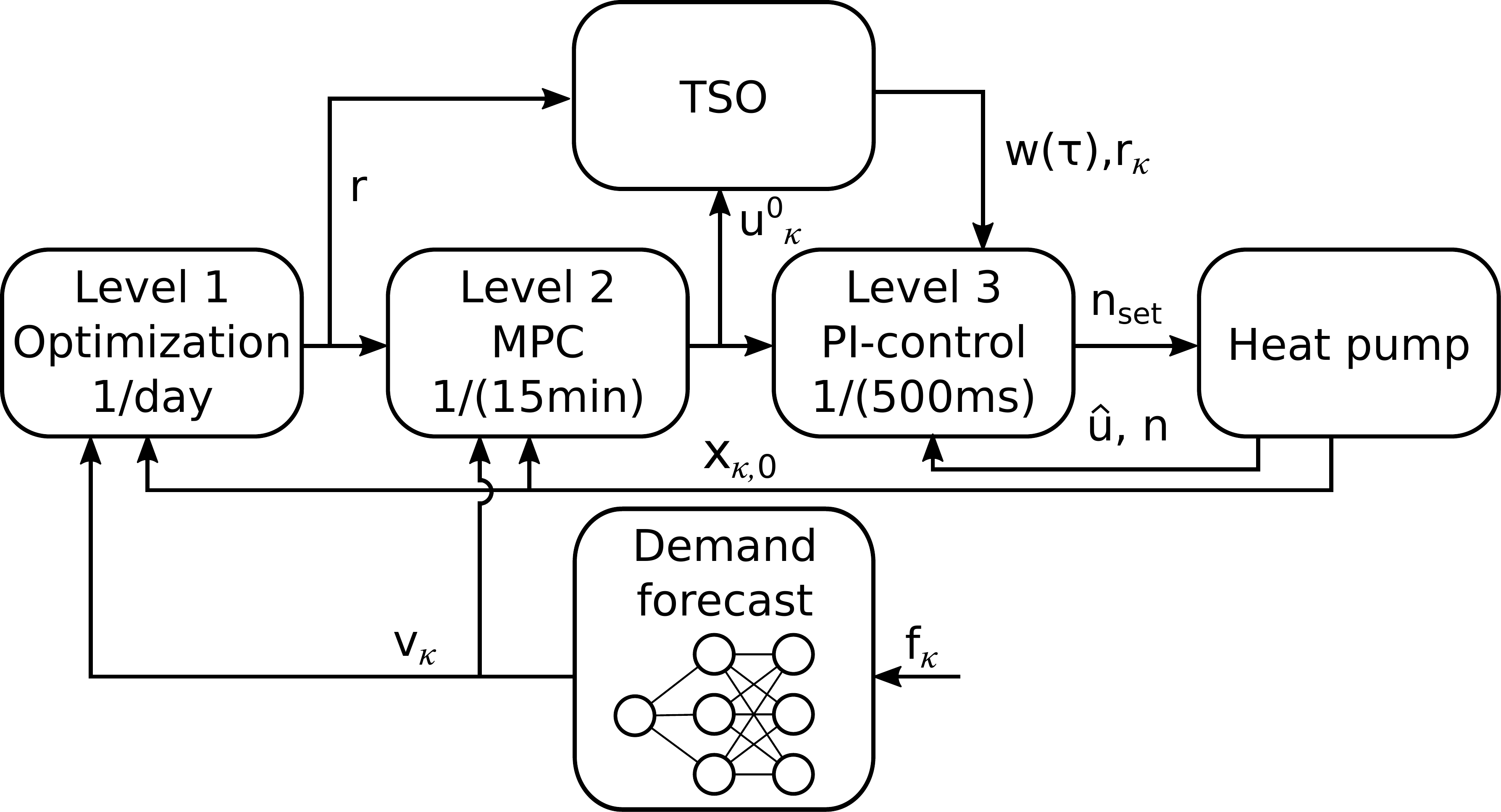}
	\caption{Hierarchical control scheme}
	\label{fig: Control scheme}
\end{figure}

To provide day-ahead frequency regulation services, we use the 3-level scheme shown in Figure \ref{fig: Control scheme}. The scheme is an extension of \cite{Vrettos2018a} and was preliminarily tested in a software-in-the-loop experiment in \cite{Bunning2020a}. Level 1 is an optimization problem solved once per day, to determine the offered reserves $r$, which are communicated to the TSO and fixed for the day ahead. Level 2 is an MPC controller, which re-optimizes the base load $u_{\kappa}^0$ of the heat pump every fifteen minutes and reacts to unforeseen disturbances. Level 3 is a lower-level feedback controller which tracks the regulated load with a fast sampling time of 0.5 seconds. Levels 1 and 2 use a heating demand forecast as an input, which is obtained from an online-corrected ANN. The required models for levels 1 and 2, along with the detailed control scheme, and the demand forecast are presented in the following section.

\section{Methodology}
\label{sec: Methodology}

\subsection{Models}
\label{subsec: Models}

The heat pump and water storage are modeled with first principles (physics based), while the heating demand of the building is modeled with the help of online corrected ANN. This is done as heat pumps and storage tanks are mass-produced industrial products for which first principles models are relatively easy to develop, while buildings are generally different from each another and thus modeling the building demand with first principles would require significant effort for each building. 

\subsubsection{Heat pump and storage model}

The heat pump, depicted on the left of Figure \ref{fig: Schematic of the heating system with heat pump, water storage tank and heat exchanger}, generates high temperature heat $u_\text{th}(t)$ by using electricity $u(t)$ and ambient heat at a lower temperature level. Here, $t$ denotes continuous time. The conversion efficiency between electrical energy and high temperature thermal energy is described with the coefficient of performance (COP) $\alpha_\text{COP}$:

\begin{equation}
u_\text{th}(t)=\alpha_\text{COP} \: u(t) + e(t).
\label{eq: heatpump}
\end{equation}

\noindent To keep the optimization problem linear, a constant $\alpha_\text{COP}$ is a reasonable assumption for ground-sourced heat pumps: as the temperature of the ground changes slowly and the heating supply temperature on the building side is usually fixed for a day (following a \textit{heating curve} dependent on the ambient temperature of the previous day), the COP can be updated, for example, on the basis of previous days' measurements and kept constant for a day. For air-sourced heat pumps, $\alpha_\text{COP}$ can be modeled as a function of the ambient temperature, which in practice can be obtained from the weather forecast. Note, however, that weather forecast uncertainty cannot be modelled, as it could lead to multiplicative uncertainty.

The error $e(t)$ captures potential additional thermal disturbances, such as COP variations based on heat pump load and internal heat losses. It can be estimated from the COP range and the maximum heat pump power, both given by manufacturer data.

Neglecting thermal losses, because they are slow and small compared to the charging and discharging of the tank, the average temperature $x(t)$ of the storage tank in Figure \ref{fig: Schematic of the heating system with heat pump, water storage tank and heat exchanger} is described by the energy balance

\begin{equation}
m \: c_p \frac{\text{d} x(t)}{\text{d} t}= u_\text{th}(t) - v(t)+\delta (t),
\label{eq: storage}
\end{equation}

\noindent where $m$ and $c_p$ denote mass and specific heat capacity of the water respectively, $v(t)$ denotes the heating demand of the building, and $\delta (t)$ denotes the error between the forecast and the actual heating demand. 
%Like the error $e(t)$ in \eqref{eq: heatpump}, $\delta (t)$ will be modeled as a box-constrained uncertainty set for the robust optimization. 
Allowing mixing of different water layers in the storage, but assuming no swapping of temperature layers, the average temperature constitutes a lower bound for the water temperature in the top layer and an upper bound for the temperature in the lowest layer, which is sufficient for our control purpose. 
While this approach might limit the exploitation of the storage tank in our control approach, it also adds robustness.
The results of the experiment will confirm that the assumptions are reasonable. Moreover, model inaccuracies compared to a stratified tank model can also be captured by $\delta (t)$. Inserting equations \eqref{eq: reserve} and \eqref{eq: heatpump} into equation \eqref{eq: storage} gives rise to the full linear description of the storage temperature:

\begin{equation}
m \: c_p \frac{\text{d} x(t)}{\text{d} t}= \alpha_\text{COP}(u^0_{k}+w(\tau) r_k) + e(t) - v(t)+\delta (t).
\label{eq: storage2}
\end{equation}

\subsubsection{Building energy demand model}

The ANN forecasting approach with online correction methods for forecasting heating demands of buildings and districts has been presented in \cite{BunningFelixBollingerAndrewHeerPhilippSmithRoy2018}. There, it was shown that the approach significantly reduces the variance in the prediction performance of the ANN, while it also increases accuracy.
%; in the study, the interquartile range of 100 different ANN reduces from 0.038 to 0.008, when correction methods are applied, while the average coefficient of determination improves from 0.818 to 0.885 in a real-life case study.  
For the sake of completeness, we reintroduce the methods here and adapt them to the forecasting task.

\begin{figure}
%	\capstart
	\centering
		\includegraphics[width=0.45\textwidth]{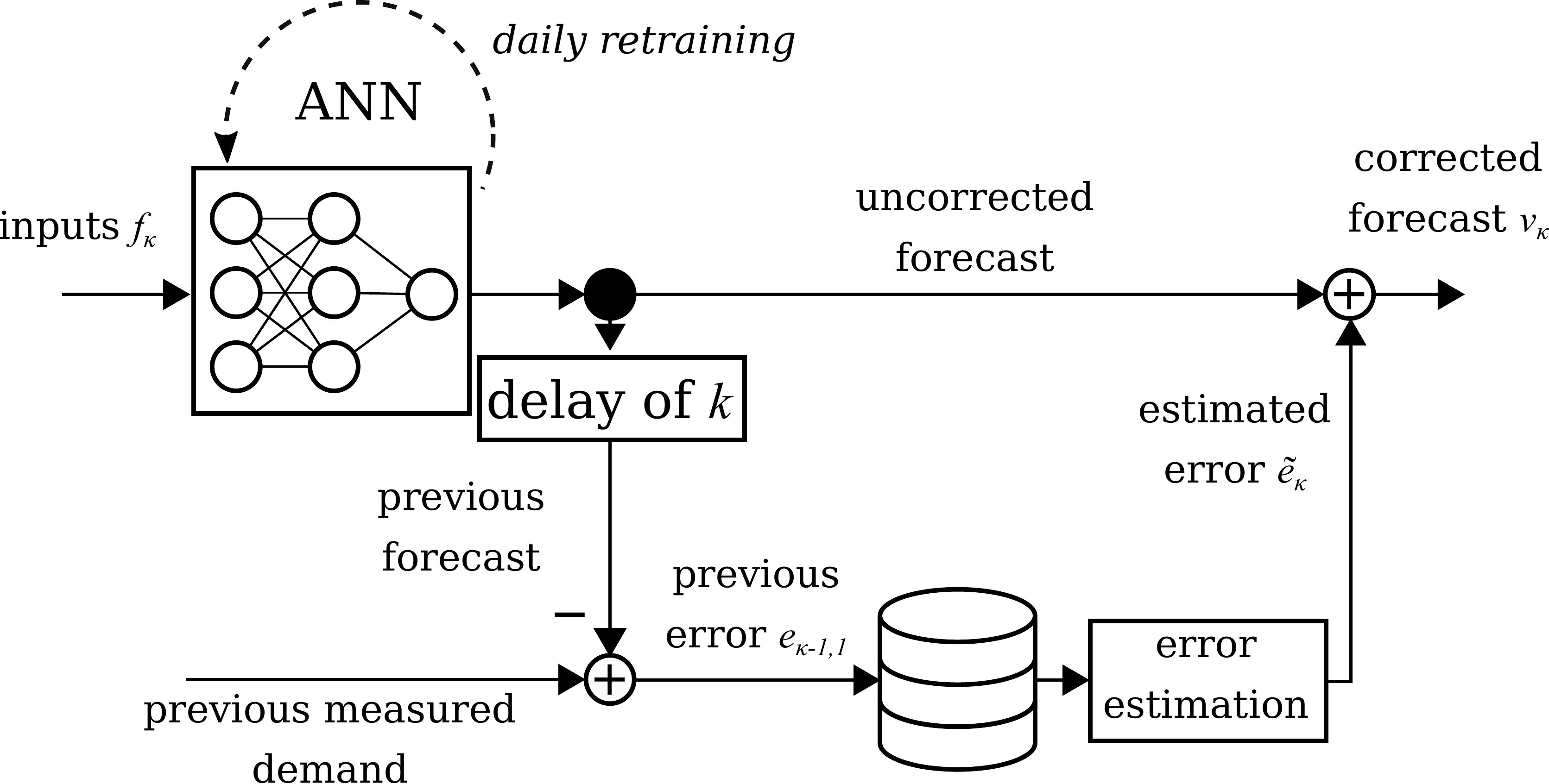}
	\caption{Forecast correction on error-autocorrelation and online learning}
	\label{fig: Forecast correction based on error-autocorrelation and online learning}
\end{figure}

For the purposes of frequency reserve provision, a heating demand forecast for a building for the next 24 hours is made starting at midnight and afterwards every 15 minutes until the end of the day. The forecasting horizon thus decreases by 15 minutes with every forecast. Both training and validation data are assumed to be sampled at 15 minute time steps. The forecast is made with a feed-forward ANN, with inputs related to ambient conditions and time features. Two correction methods are applied in the online phase of the forecasting task (Figure \ref{fig: Forecast correction based on error-autocorrelation and online learning}).

The first correction method is based on the forecasting \textbf{error-autocorrelation}. The error $\tilde{e}$ of the forecast conducted at the current time $\kappa$ for forecasting interval $k$ is estimated with

\begin{equation}
\tilde{e}_{\kappa,k}=e_{\kappa-1,1} R_{ee}(k,\mathcal{E})
\label{eq: error correction1}
\end{equation}

\noindent where

\begin{equation}
R_{ee}(l,\mathcal{E})=\frac{\mathbb{E}[(\mathcal{E}-\mu)(\mathcal{E}_{+l}-\mu)]}{{\sigma}^2}.
\label{eq: error correction2}
\end{equation}

\noindent Here, $e_{\kappa-1,1}$ denotes the difference between the first (15-minute) element of the last conducted forecast (at time $\kappa-1$) and the actual measured heating demand. $R_{ee}(l,\mathcal{E})$ is the autocorrelation of the forecasting error, which is dependent on a time-lag $l$ and the set of all past forecasting errors $\mathcal{E}$, including the training and testing data sets as well as the data gathered during online operation. $\mathcal{E}_{+l}$ is the corresponding set shifted in time by $l$. The properties of the underlying stochastic process (expected value $\mathbb{E}$, mean $\mu$ and standard deviation $\sigma$) are empirically approximated based on the set $\mathcal{E}$.

The rationale behind the correction is based on the assumption that forecasting errors persist over time because the source for these errors also persist over time in a building; For example, opening a window will likely have an impact on the heating demand for a longer time period than a single 15-minute interval. The last measured forecasting error can thus be used to correct the next forecast.

The correction procedure is illustrated in Figure \ref{fig: Forecast correction based on error-autocorrelation and online learning}. A forecast is made at time $\kappa$ based on the inputs $f_{\kappa}$. The previous forecast from time $\kappa-1$ is compared to the actual measured heating demand, giving rise to $e_{\kappa-1,1}$. With all previously measured errors $\mathcal{E}$, stored in a database, $\tilde{e}_{\kappa}$ can be calculated using equations (\ref{eq: error correction1}) and (\ref{eq: error correction2}). Adding $\tilde{e}_{\kappa}$ to the uncorrected forecast gives rise to the corrected forecast $v_{\kappa}$, which will later be used as an input to the control scheme.

The second forecasting correction method is based on \textbf{online learning}: Instead of only training the ANN on a training set offline and using the ANN for predictions online, the ANN is retrained online every 24 hours on the basis of the data gathered during the previous day. (This is symbolized by the dotted loop in Figure \ref{fig: Forecast correction based on error-autocorrelation and online learning}). By doing this, changes to the building that persist for longer than one day can be captured; such changes could include changing the set point of a thermostat for example.

While conventional wisdom suggests that ANN require large data sets for training, due to the correction methods, the presented approach can already reach high prediction performance with a single week of training data \cite{Bunning2019a}. For further details on the method and numerical results we refer to \cite{BunningFelixBollingerAndrewHeerPhilippSmithRoy2018}.

\subsection{Controller design}
\label{subsec: Controller design}

The models and demand forecast developed in Section \ref{subsec: Models} are used in the 3-level control scheme depicted in Figure \ref{fig: Control scheme}. Level 1 solves a robust optimization problem once every 24 hours at midnight. Based on the current storage tank temperature $x_{\kappa,0}$ and the heating demand forecast of the building $v_\kappa$ it determines the reserves $r$ to be offered and fixed in 15 minute intervals, $r_k$, for the next 24 hours. The index $\kappa \in [1,96]$ denotes the discrete time index, e.g. $\kappa = 1$ for midnight and $\kappa = 2$ for 00.15 a.m., while $k \in [1,N]$ denotes the index in the optimizations: for example, $u^0_{3,8}$ is the eighth element of the heat pump base consumption in the optimization conducted at time $\kappa=3$ (00.30 a.m.).

Level 2 solves an optimization problem similar to the one in Level 1 every 15 minutes during the day, with a shrinking horizon, from the current time to midnight. In this optimization problem, the values of the reserves $r$ for the rest of the day are already known, because they have been fixed by Level 1. Further inputs to this problem are the updated tank temperature measurement $x_{\kappa,0}$ and the updated heating demand prediction $v_{\kappa}$. The outputs of Level 2 are the nominal heat pump electrical power set points $u^0_{\kappa}$ for each 15 minute interval for the remainder of the day, of which the first one, $u^0_{\kappa,1}$, is passed on to Level 3. 

Level 3 is a Proportional-Integral controller that controls the relative rotational speed $n$ of the heat pump's compressor to track the regulated heat pump's electricity consumption $u_{\kappa}(\tau)$.

\subsubsection{Level 1}

For Level 1, equation \eqref{eq: storage} can be rearranged to the continuous state space system

\begin{equation}
\begin{aligned}
\frac{\text{d} x(t)}{\text{d} t} & = \frac{1}{m \: c_p } \left( u_\text{th}(t) - v(t)+\delta (t) \right)\\ & = A_{cont} x + B_{cont} \left( u_\text{th}(t) - v(t)+\delta (t) \right),
\end{aligned}
\label{eq: storage3}
\end{equation}

\noindent with $ A_{cont}$ and $B_{cont}$ denoting the continuous state and input matrix respectively. Discretizing in time, and redefining $x :=[x_1,...,x_N]^\top \in \mathbb{R}^N$, $u :=[u_1,...,u_N]^\top \in \mathbb{R}^N$, $v:=[v_1,...,v_N]^\top \in \mathbb{R}^N$, $\delta:=[\delta_1,...,\delta_N]^\top \in \mathbb{R}^N$, with $\top$ denoting a transpose, and $N$ denoting the prediction horizon, we describe the discrete state trajectory by

\begin{equation}
x=A x_0+B(u_{th}-v+\delta),
\label{eq: LTI1}
\end{equation}

\noindent where $x_0$ is the initial state of the system, and $A$ and $B$ are defined appropriately \citep{Borrelli2017}. As a discretization method, both the exact method and Euler method can be used: as \eqref{eq: storage} does not model storage losses (which are negligible compared to the charging and discharging of the tank by the heat pump and connected buildings), both continous and discrete $A$ matrices are equal to 1 for any discretization method. Potential errors arise instead from the implicit assumption that the inputs are constant over the sampling period. Therefore, the sampling time is chosen according to common practices in the building control domain \citep{schwenzer2021review}. It is also related to the reserve scheme defined by \eqref{eq: reserve}.

By vectorizing all remaining variables and adding equation \eqref{eq: heatpump} to the formulation, the robust optimization problem in terms of the offered reserves $r$, the nominal heat pump electrical set points $u^0$, and the heat pump on/off condition $z$, can be written as

\begin{comment}
Discretizing in time using Euler discretization, leads to the discrete-time state space model

\begin{equation}
x_{k+1}=\tilde{A}x_k+\tilde{B}(u_{th,k}-v_k+{\delta}_k).
\label{eq: LTI1}
\end{equation}

\noindent By defining 

\begin{equation}
A:=
\begin{pmatrix}
\tilde{A} \\
\tilde{A}^2 \\
\vdots \\
\tilde{A}^N \\
\end{pmatrix}
\quad
B:=
\begin{pmatrix}
\tilde{B} & 0 & \cdots & 0 \\
\tilde{A}\tilde{B} & \tilde{B} & \ddots & 0 \\
\vdots & \ddots & \ddots & \vdots \\
\tilde{A}^{N-1}\tilde{B} & \cdots & \tilde{A}\tilde{B} & \tilde{B}
\end{pmatrix}
,
\end{equation}

\noindent where $N$ denotes the horizon, and redefining $x :=[x_1,...,x_N]^\top \in \mathbb{R}^N$, $u :=[u_1,...,u_N]^\top \in \mathbb{R}^N$, $v:=[v_1,...,v_N]^\top \in \mathbb{R}^N$, $\delta:=[\delta_1,...,\delta_N]^\top \in \mathbb{R}^N$, with $\top$ denoting a transposition, we describe the state trajectory by

 \begin{equation}
x=A x_0+B(u_{th}-v+\delta),
\label{eq: LTI1}
\end{equation}

\noindent where $x_0$ is the initial state of the system.
\end{comment}

\begin{subequations}
\begin{alignat}{2}
&\!\min_{r,x,u^0,u_\text{th},\epsilon ,z}         &\qquad& {f^\text{el}}^\top u^0 - {f^{r}}^\top r + \lambda ^\top \epsilon \label{eq:optProb}\\
&\text{subject to} &      & x = A x_0 + B (u_\text{th}-v+\delta+e),\label{eq:constraint1}\\
&                  &      & u_\text{th}=\alpha_\text{COP}(u^0+w\odot r)\label{eq:constraint2},\\
&				   & 	  & X_\text{min} - \epsilon \leq x \leq X_\text{max} + \epsilon\label{eq:constraint3}, \\
&				   &      & z U_\text{min} \leq u^0+ w\odot r \leq z U_\text{max}\label{eq:constraint4}, \\
&				   &      & z \in \mathbb{Z}^N_2 , \\
&				   &	  & \epsilon \geq 0, \\
&				   &	  & \forall w\in W, \forall \delta \in \Delta, \forall e \in E. \label{eq:qual1}
\end{alignat}
\end{subequations}

\noindent Here, $f^\text{el}$ and $f^{r}$ denote costs for electricity and benefits for offered reserves respectively. $X_\text{min}$ and $X_\text{max}$ describe temperature limits for the storage tank, defined by the lowest possible operating temperature for floor heating, which ensures the indoor thermal comfort, and the highest supply temperature of the heat pump. The slack variable $\epsilon \in \mathbb{R}^N$ ensures feasibility with respect to the storage temperature constraint and $\lambda$ denotes the associated cost. The lower and upper electrical capacity limits of the heat pump are described by $U_\text{min}$ and $U_\text{max}$, and $z \in \mathbb{Z}^N_2$ is a binary variable that determines if the heat pump is switched on or off. The symbol $\odot$ denotes the operator for element-wise multiplication. All constraints have to hold for all realizations of uncertainties $w\in W, \delta \in \Delta, e \in E$. Note that the heat pump error $e$ could further be coupled to the heat pump on-off status to avoid accounting for unrealistic error realizations. Suitable sets for $W$ can be found by analysing the regulation signal $w(\tau)$. Note that $r$ will always be positive due to the formulation of the cost function. However, as $W$ can contain negative regulation signals, the effective reserve can become negative.

Constraints (9c) and (9e) can be reformulated as linear constraints by making $w$ a square diagonal matrix. Thus, problem (9) is a Mixed Integer Linear Program (MILP) that has to hold for the qualifier \eqref{eq:qual1}. While $W$, $\Delta$ and $E$ generally allow any convex sets, for box-constrained sets the robust optimization problem (9) can be reformulated as a Mixed Integer Linear Program via \textit{explicit maximization} \cite{Lofberg2012}. This reformulation is performed automatically by many modern optimization tools \citep{Lofberg2004,Goh2011}.

The binary variable $z$ forces $u^0$ and $r$, and thus $u_{th}$, to be zero if the electrical input to the heat pump does not exceed $U_{min}$. For the heating system this means that in case of low heating demand from the building, a hysteresis behaviour can be expected, where the heat pump changes between over-serving the demand and switching off. Potentially, there could be combinations of $X_{min}$, $X_{max}$, $U_{min}$ and $U_{max}$ where the heating demand could not be served, but this issue is captured by the slack variable $\epsilon$.  

In the case of low storage capacities, i.e. large $B$, corresponding to low mass of water, low $X_\text{max}$ or high $X_\text{min}$, or large uncertainty in $W$, $\Delta$, and $E$, the offered reserves $r$ may become very small or, without the slack variable $\epsilon$, the problem may even become infeasible. This is because the uncertainty induced in $x$ by the action of $w$, $\delta$, and $e$ compounds along the horizon, as uncertainty at subsequent steps gets added to that of earlier steps through the “integrator” implicit in \eqref{eq:constraint1} (see \eqref{eq: storage}). As a consequence, near the end of the horizon the uncertainty in x becomes large, leading to a violation of \eqref{eq:constraint3}. This growth in uncertainty traces its origins to the fact that (9) addresses Level 1 in the control hierarchy of Figure 3, but does not contain any information about the actions of the lower levels. In reality, Level 2 and Level 3 will be executed repeatedly within the horizon of (9), adjusting the decisions of Level 1 to account for information that has become available in the meantime. This introduces feedback to the process, that will in practice limit the growth of the uncertainty. 

In stochastic programming, information about this “recourse” process can be introduced by optimising over causal feedback policies instead of a sequence of “open-loop” decisions fixed at the beginning of the horizon. In this case, the optimization problem for Level 1 encodes the fact that the system will react to uncertainties that are still unknown at the time (9) is solved, but will been revealed at the time the decision is implemented. As discussed in \cite{Warrington2012}, optimizing over the set of all possible policies is intractable in general. To obtain a tractable optimization problem, one can restrict the classes of causal policies considered. A common choice in this respect is the class of affine disturbance policies \cite{Ben-Tal2004,Goulart2006}. For the uncertainties introduced by the regulation signal $w$, equation (\ref{eq:constraint2}) can be extended to

\begin{equation}
u_\text{th}=\alpha_\text{COP}(u^0+w\odot r + D_w w),
\end{equation}

\noindent where $D_w \in \mathbb{R} ^{NxN}$ is a strictly lower triangular matrix:

\begin{equation}
D_w:=
\begin{pmatrix}
0 			& 0 		&\cdots			& 0 \\
[D_w]_{2,0} & 0 		& \ddots 		& 0 \\
\vdots 		& \ddots 	& \ddots 		& 0 \\
[D_w]_{N,0} & \cdots 	& [D_w]_{N,N-1} & 0
\end{pmatrix}
.
\end{equation}

\noindent By making $D_w$ a decision variable in the optimization problem, the uncertainty in $u_{th}$ can be lowered, and thus also the uncertainty in $x$. Affine policies on the other uncertain variables $\delta$ and $e$ can also be defined. Because $\delta$ and $e$ appear together in (\ref{eq:constraint1}), a single lower triangular matrix can be used:

\begin{equation}
u_\text{th}=\alpha_\text{COP}(u^0+w\odot r + D_w w + D_{\delta ,e} (\delta +e)).
\end{equation}

We note that the regulation signal that takes values in the interval [-1, 1] is updated every 2 seconds, but the rest of the decision variables in (9) refer to quantities that are updated every 15 minutes. Therefore, when trying to meet the robust constraint (9c) the average value of $w(\tau)$ over a 15 minute interval (denoted by $\bar{w}$ below) is more relevant than the instantaneous value. By collecting historical data, a second uncertainty set on the average of $w(\tau)$ can be created by integrating over 15-minute horizons and evaluating the distribution of these integrals (see \cite{Vrettos2018a}). As a result, the uncertainty set is decreased to $\bar{W} \subset W$ for constraint (\ref{eq:constraint2}). As the instantaneous electrical consumption needs to remain within operational limits at all times, $w\in W$ remains for constraint (\ref{eq:constraint4}). 

The resulting optimization problem is

\begin{subequations}
\label{prob2}
\begin{alignat}{2}
&\!\min_{\substack{r,x,u^0,u_\text{th},z,\\ \tilde{z},D_w,D_{\delta ,e},\epsilon}}         &\qquad& {f^\text{el}}^\top u^0 - {f^{r}}^\top r + \lambda ^\top \epsilon \label{eq:optProb2}\\
&\text{subject to} &      & x = A x_0 + B (u_\text{th}-v+\delta +e),\label{eq:constraint1_2}\\
&                  &      & u_\text{th}=  \alpha_\text{COP}(u^0+\bar{w}\odot r \nonumber \\
&				   &	  & \hspace{7mm} +D_w \bar{w}+D_{\delta ,e} (\delta +e))\label{eq:constraint2_2},\\
&				   & 	  & X_\text{min} - \epsilon \leq x \leq X_\text{max} + \epsilon \label{eq:constraint3_2}, \\
&				   &      & z U_\text{min} \leq u^0+w\odot r +D_w \bar{w} \nonumber \\
&				   &      & \hspace{10mm} +D_{\delta ,e} (\delta +e) \leq z U_\text{max}, \label{eq:constraint4_2}\\
&			    	   &      & \tilde{z}R_{min} \leq r \leq \tilde{z}R_{max}, \label{eq:constraint5_2}\\
&				   &      & z, \tilde{z} \in \mathbb{Z}^N_2 , \\
&				   &	  & \epsilon \geq 0, \\
&				   &      & [D_w]_{i,j} = 0 \: \forall j \geq i, \\
&				   &      & [D_{\delta ,e}]_{i,j} = 0 \: \forall j \geq i, \\
&				   &	  & \forall w\in W, \forall \bar{w}\in \bar{W}, \nonumber \\
&				   &	  & \forall \delta \in \Delta, \forall e \in E.
\end{alignat}
\end{subequations}

\noindent The heat pump capacity constraint, now (\ref{eq:constraint4_2}), is adapted to ensure feasibility under the chosen policies. Note that, $[D_{\delta ,e}]_{k,j}$ and $[D_w]_{k,j}$ (as well as $u^0_k$ and $r_k$) will be zero whenever $z_k=0$. Moreover, as the results of \cite{Bunning2020a} suggested that small reserves $r$ lead to weak tracking performance (and low performance scores) because of large relative errors, a second binary variable $\tilde{z}$ was added to impose a lower limit on $r$ through constraint (\ref{eq:constraint5_2}). Problem (13) is still a MILP.

\subsubsection{Level 2}

Controller Level 2 is a MPC scheme with shrinking horizon. The purpose of Level 2 is to introduce feedback, i.e. change the planned control inputs due to newly available information. The horizon is shrinking, because $r$ is only fixed by Level 1 until the end of the day. This choice could potentially lead to storage depletion towards the end of the day if no reserves are offered during these intervals. However, as we use a slack variable on the state constraints, there are no issues with recursive feasibility.\footnote{Moreover, as the system is stable and in practice often operated with hysteresis control, even without slack variables, recursive feasibility is ensured.} At most, storage depletion will lead to low reserves being offered in the first intervals of the following day. Alternative formulations include a receding horizon scheme, which allows Level 2 to choose those reserves $r$ at the end of the horizon that are not yet fixed by Level 1. However, this significantly increases the computational complexity.

Level 2 can update $u^0$, depending on new measurements of initial conditions $x_0$ and demand forecasts $v$, which are updated every 15 minutes. It uses the same optimization problem as Level 1, with the exception that $r$ is already fixed from Level 1 for the whole day, and is thus not a decision variable any more in Level 2. Other than $r$, no information from Level 1 is carried over to Level 2. 

\begin{comment}
\begin{subequations}
\label{prob2}
\begin{alignat}{2}
&\!\min_{\substack{x,u^0,u_\text{th},z, \\ D_w,D_{\delta ,e},\epsilon}}         &\qquad& {f^\text{el}}^\top u^0 + \lambda ^\top \epsilon \label{eq:optProb3}\\
&\text{subject to} &      & x = A x_0 + B (u_\text{th}-v+\delta +e),\label{eq:constraint1_3}\\
&                  &      & u_\text{th}=  \alpha_\text{COP}(u^0+\bar{w}\odot r \nonumber \\
&				   &	  & \hspace{7mm} +D_w \bar{w}+D_{\delta ,e} (\delta +e))\label{eq:constraint2_3},\\
&				   & 	  & X_\text{min} - \epsilon \leq x \leq X_\text{max} + \epsilon \label{eq:constraint3_3}, \\
&				   &      & z U_\text{min} \leq u^0+w\odot r +D_w \bar{w} \nonumber \\
&				   &      & \hspace{10mm} +D_{\delta ,e} (\delta +e) \leq z U_\text{max}, \label{eq:constraint4_3}\\
&				   &      & z \in \mathbb{Z}^N_2 , \\
&				   &	  & \epsilon \geq 0, \\
&				   &      & [D_w]_{i,j} = 0 \: \forall j \geq i, \\
&				   &      & [D_{\delta ,e}]_{i,j} = 0 \: \forall j \geq i, \\
&				   &	  & \forall w\in W, \forall \bar{w}\in \bar{W}, \nonumber \\
&				   &	  & \forall \delta \in \Delta, \forall e \in E.
\end{alignat}
\end{subequations}
\end{comment}

The optimization schemes in Levels 1 and 2 do not consider electricity peak pricing. However, it can be easily included in the scheme as described in \citep{Zhang2015,Sundstrom2017,Ma2014a}. Moreover, although we use a fixed electricity price in the case studies, $f^\text{el}$ can in principle be variable; for example, time-of-use pricing can also be modeled.

\subsubsection{Level 3}

Level 3 is a discrete Proportional-Integral feedback controller, with proportional gain $k_p$ and integral gain $k_i$, to track equation \eqref{eq: reserve} with the heat pump. The controller output is the set point for the relative rotational compressor speed of the heat pump $n_\text{set}$. The controller input is the heat pump's measured electrical load $\hat{u}$. An anti-windup scheme is used in case the heat pump reaches its compressor speed limitations. The integration block of the controller is also bypassed if the difference between the set compressor speed $n_\text{set}$ and the measured compressor speed $n$ exceeds a limit. This is done because heat pumps usually have up and down ramping limits.

\section{Experimental case study and results}
\label{sec: Experimental case study and results}

\subsection{Configuration}

\begin{figure}
%	\capstart
	\centering
		\includegraphics[width=0.45\textwidth]{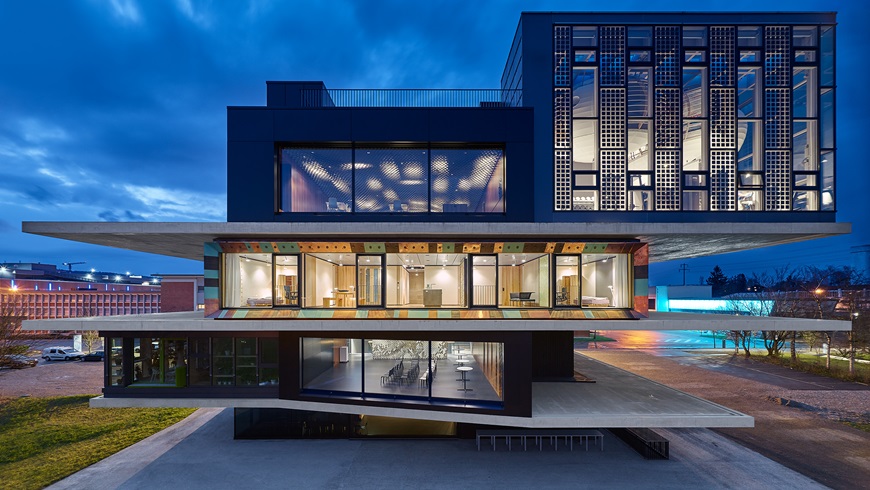}
	\caption{NEST demonstrator at Empa in Switzerland. Open areas can accommodate future experimental units. Copyright: Zooey Braun - Stuttgart}
	\label{fig: NEST building at Empa in Switzerland, Copyright: Zooey Braun - Stuttgart}
\end{figure}

\begin{figure}
%	\capstart
	\centering
		\includegraphics[height=4cm]{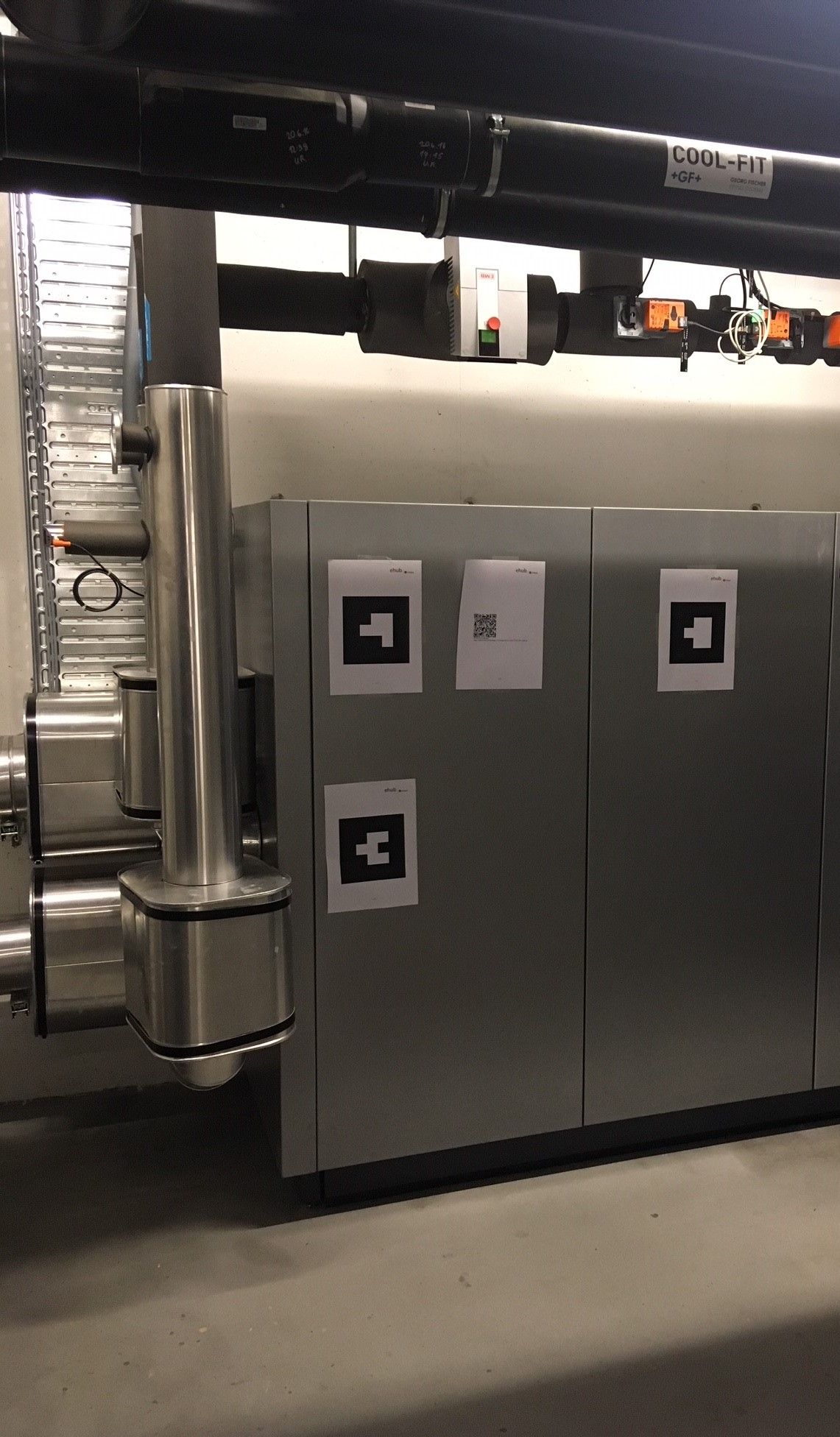}
		\includegraphics[height=4cm]{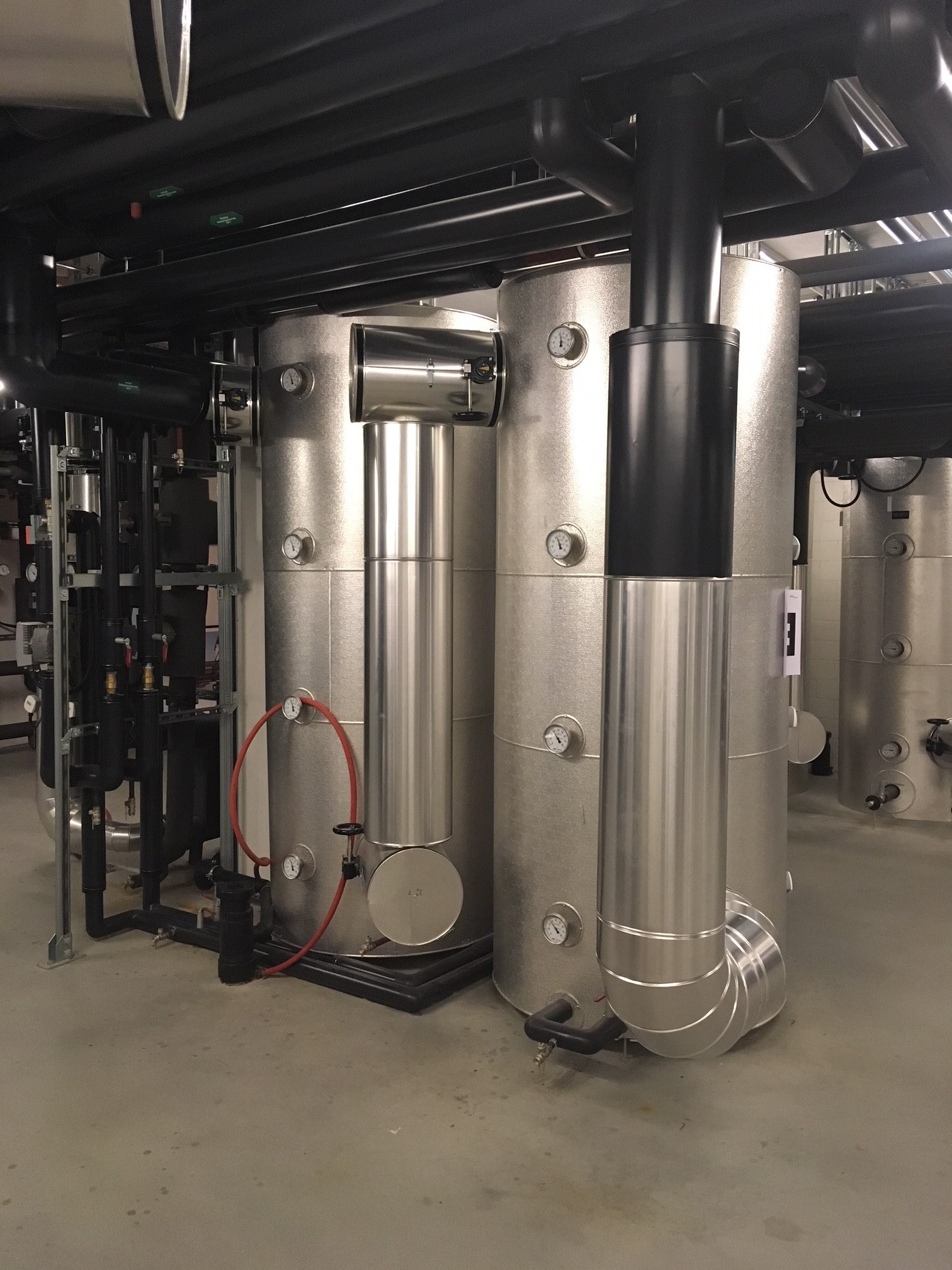}
	\caption{Experimental facility with heat pump and water storage tanks \cite{Bunning2020a}}
	\label{fig: Experimental set-up with heat pump and water storage tanks}
\end{figure}

We test the reserve scheme in a three-day experiment on a real system in the NEST building (Figure \ref{fig: NEST building at Empa in Switzerland, Copyright: Zooey Braun - Stuttgart}) at Empa, Switzerland. The building consists of individual residential, office and multi-use units that can be added and removed from the building backbone, as well as permanent office and meeting rooms. The individual units are connected to a central heating system with a supply temperature of 38 {\degree}C and a return temperature of 28 {\degree}C via heat exchangers and are equipped with their independent control systems considered to be unknown in the experiment. Due to the independence of the connected units and their controllers, and the structure of the heating system, the building is often referred to as a ``vertical district'' \citep{Empa} or ``vertical neighborhood'' \citep{Richner2018} and is intended to serve as an emulator for studies on concepts considered for district heating systems. 

The heating system (Figure \ref{fig: Experimental set-up with heat pump and water storage tanks}) comprises a ground source heat pump, specifically the two-compressor model WP-WW-2NES 20.F4-2-1-S-P100 produced by Viessmann with a maximum thermal capacity of 100 kW, and a water buffer storage consisting of two  1100 litre Matica water tanks connected in series. Only the first compressor stage is used in the experiment, which leads to a rated maximum thermal capacity of 50 kW. This is done, as the the power of the connected ground borehole field is not sufficient to supply both compressors. The system resembles the configuration described in Section \ref{sec: Problem statement}.

\begin{table}[]
	\caption{Parameters for controller Levels 1 and 2}
	\label{tab: Parameters for controller level 1 and 2}
	\centering
	\footnotesize
	\begin{tabular}{|l|l|l|}
	
	\hline

	 $N=96$,   					& $\alpha_\text{COP}=3.53$,  				& $\tilde{A}=1$,  \\
	 $\lambda=5$, 			& $W=[-1,1]$, 					& $\tilde{B}=0.0978 \frac{K}{kW},$  \\
	 $f^\text{el}=1$,  				& $\bar{W}=[-0.25,0.25]$, 							& $X_\text{min}=28 \degree C$,    \\
	 $f^\text{r}=1.5$ 				& $U_\text{min}=8.2 kW$, 				& $X_\text{max}=38 \degree C$, \\ 
	 $R_\text{min}=0.4 kW$		& $U_\text{max}=12.8 kW$,						& $E \oplus \Delta=[-4.0,4.0] \: kW$ \\ \hline
	
	\end{tabular}

\end{table}

The control scheme configuration is as follows. The parameters for controller Levels 1 and 2 are shown in Table \ref{tab: Parameters for controller level 1 and 2}. 
The cost-function related parameters $\lambda$, $f^\text{el}$, and $f^\text{r}$ were chosen based on preliminary numerical studies to balance the trade-off between cost optimality and constraint violation. Compared to \cite{Bunning2020a}, we have increased the reserve benefit $f^\text{r}$ to get richer $r$ vectors to test the robustness of the controller. They are not directly related to a specific energy product and therefore considered dimensionless. The values for $\alpha_\text{COP}$, $U_\text{min}$, $U_\text{max}$ and $R_{min}$ were set on the basis of preliminary heat pump experiments. We choose a constant COP as the heat pump is ground-sourced and supply and return temperatures are relatively constant on the building side. The limits for $W$ are properties of the used regulation signal RegD by PJM. The uncertainty set $\bar{W}$ can be determined by analyzing historical regulation signals \cite{Vrettos2018a}, $\tilde{A}$ follows from the assumption of no thermal losses, and $\tilde{B}$ is calculated on the basis of the tank volume and the specific heat capacity of water. In contrast to \cite{Bunning2020a}, where we chose the set boundaries based on historical measurement data from the building and the heat pump, we do not specify the uncertainty sets $E$ (error from disturbances of the heat pump) and $\Delta$ (error from demand forecast) separately, but instead define the Minkowski sum $E \oplus \Delta$, and shrink it compared to the original source. The initial value of the average tank temperature, $x_0$, is determined by taking a weighted average of six temperature sensor measurements at different heights within the storage tank. To reduce wear on the heat pump, we introduce an additional constraint to Level 1, and require the on/off status of the heat pump $z_k$ to be constant during each thirty minute interval.

%\footnote{The shrunk uncertainty set is a result of the robust experimental results in \cite{Bunning2020a} but there is also a rough theoretical justification: The variance, $\sigma_{X+Y}$, of the sum of two uncorrelated normal distributions $X$ and $Y$ is smaller than the sum of the variance of the individual distributions, namely $\sigma_{X+Y}=\sqrt{\sigma_{X}^2+\sigma_{Y}^2}$.}

The proportional and integral gains, $k_p=2.0$ and $k_i=0.4$, of the PI controller in Level 3 were determined by first modeling and auto-tuning a first-order representation of the heat pump in Simulink\textsuperscript{\textregistered}, and then manually adjusting the values after implementing the controller on the actual plant. The controller output limits for $n_\text{set}$ are set to 20\% and 50\% of the relative compressor speed, as the second compressor stage of the heat pump is activated if the relative speed exceeds the limit of 50\%. The controller sampling time is 500 milliseconds and the maximum allowed difference between controller output $n_\text{set}$ and measured compressor speed $n$ before anti-windup activates is set to 2\%. Moreover, as the heat pump is only fully controllable five minutes after switching it on, it is switched on five minutes early in the case that reserves are offered for the next 15-minute interval.

For the implementation of Levels 1 and 2 we use Matlab\textsuperscript{\textregistered}. As all uncertain variables are box-constrained, the optimization problems become MILPs. These are written with YALMIP \cite{Lofberg2012}, which automatically derives robust counterparts, and are solved with CPLEX\textsuperscript{\textregistered} 12.9. Each optimization is started five minutes before the decision is implemented, limiting the solver time to five minutes. This time is enough to solve the problem close to optimality in all cases.\footnote{We conducted preliminary numerical studies to investigate how the solution converges with time.} The best feasible solution is then implemented. The Level 3-controller is written in Python 3. The communication of the optimization results and the sensor measurements between Matlab and Python is facilitated via shared csv files. A Python OPC-UA client is used for the communication with sensors and actuators of the heat pump and the building. The heat pump takes as an input the relative compressor speed (as discussed above) and uses an internal controller to track it.

The heating demand forecast is performed at midnight (for Level 1) and then every 15 minutes (for Level 2) with an ANN and the correction methods presented in Section \ref{sec: Methodology}. The correction based on error-autocorrelation is applied with every new forecast, while the online retraining is done only at midnight. The ANN model uses as inputs the forecast ambient temperature (broadcast by MeteoSwiss and updated every 12 hours), the hour of the day (which is one hot encoded), the measured heating demand one day ago at the same time, the measured heating demand one week ago at the same time, and a binary variable that indicates whether it is a working or a non-working day. The framework also allows to include inputs such as indoor temperatures, thermostat positions, and occupancy forecasts from the connected buildings. However, in this study, they are excluded to avoid privacy-related issues. The ANN model is implemented in Python 3 with Keras \cite{Chollet2018}. It is a feed-forward network with two hidden layers and 8 nodes per hidden layer with Rectified Linear Units (ReLu) as activation functions. Just short of three years of historical data (sampled in 15 minute intervals) were used for training using the optimizer \textit{adam} \cite{Yun2018} with the standard learning rate of 0.001, a batch size of 1, and 10 epochs. This configuration corresponds to the one presented and validated in \cite{BunningFelixBollingerAndrewHeerPhilippSmithRoy2018}.

The experiment was conducted on three consecutive days from the 25th of February 2020, 11.45 am, to the 28th of February, 11.45 am. As a regulation signal, the RegD signal by PJM from the 27th of January 2019 was used for all three days. This choice was made for convenience, as the signal starts and ends with a value of 1 and is thus continuous when repeated. It does not significantly affect the results as, the regulation signal is fast (see Figure \ref{fig: Results tracking performance} (b)) and the statistics of the signal are similar across different days. Moreover, signals are not correlated across days \citep{Vrettos2016thesis}, which makes any combination of signals equally likely on two consecutive days. 

\subsection{Results}

\begin{figure*}
%	\capstart
	\centering
		\includegraphics[width=1.0\textwidth]{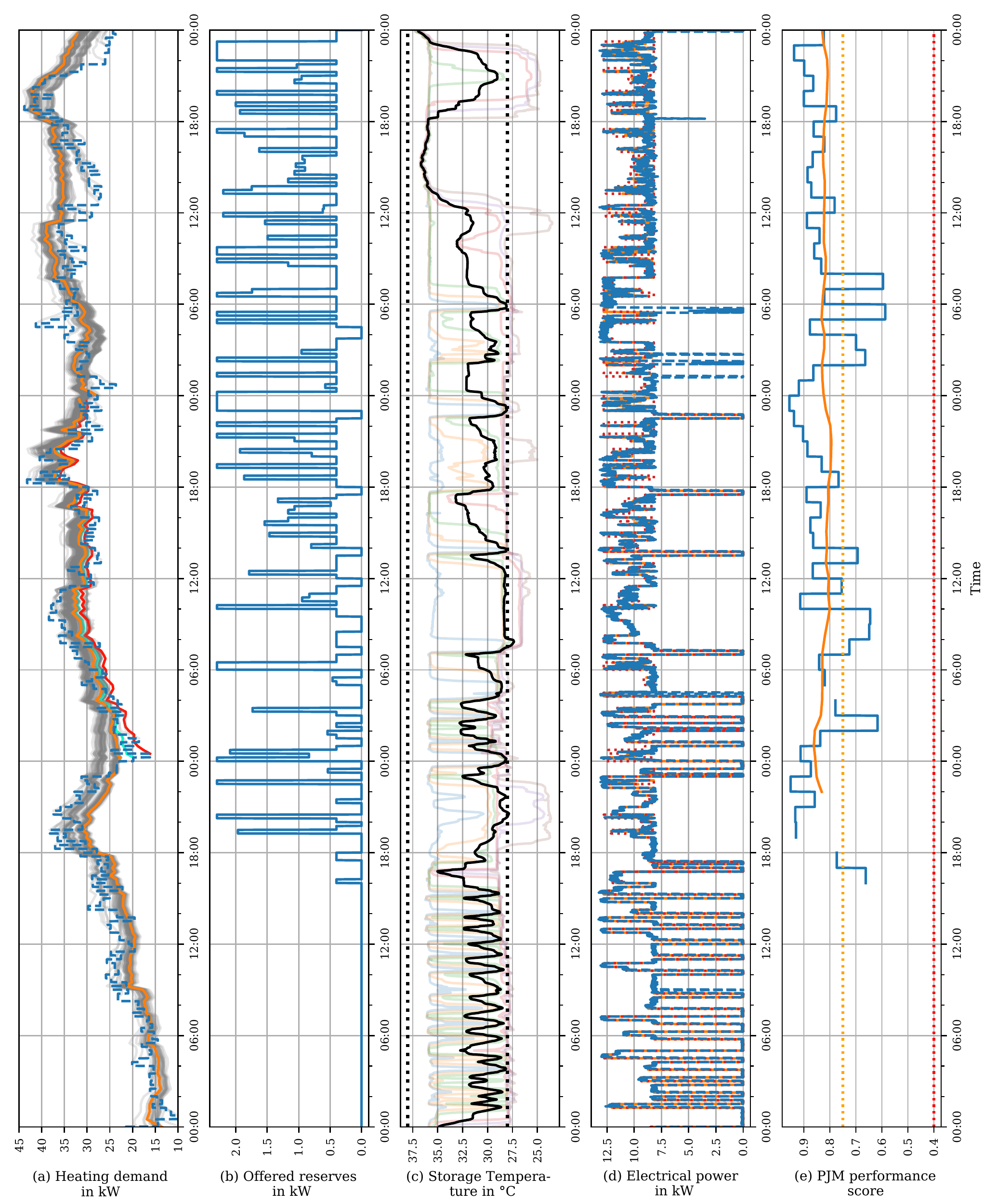}
	\caption{Complete experimental results. (a): measured heating demand in dashed blue, daily forecast in orange, 15-minute forecasts in transparent grey, specifically mentioned forecasts in cyan and red, (b): offered reserves in blue, (c): average tank temperature in black, temperature constraints in dotted black, individual layer temperatures in transparent colours, (d): set point for electrical power in orange, measured electrical reserves in dashed blue, potential power range due to regulation signal in dotted red, (e): PJM performance score in blue (no data during hours where no reserves are offered), 20-hour moving average of performance score in orange, qualification limit in dotted orange, operation limit in dotted red.}
	\label{fig: Results whole experiment}
\end{figure*}

\begin{figure*}
%	\capstart
	\centering
		\includegraphics[width=1.0\textwidth]{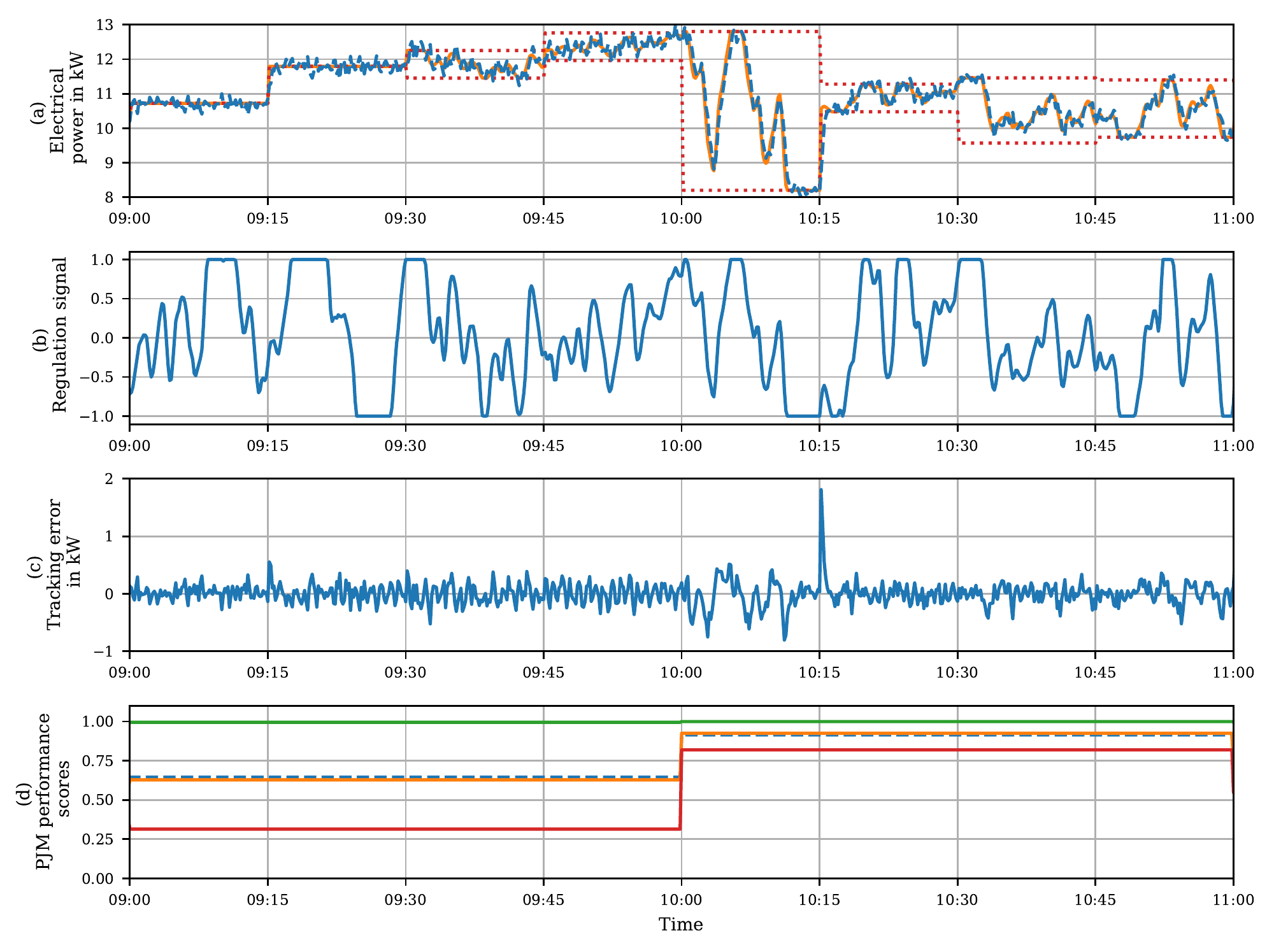}
	\vspace{-3mm}
	\caption{Results tracking performance, (a): set point for electrical power in orange, measured electrical reserves in dashed blue, potential power range due to regulation signal in dotted red, (b): regulation signal, (c): tracking error, (d): delay score in green, accuracy score in orange, precision score in red, composite score in dotted blue.}
	\label{fig: Results tracking performance}
\end{figure*}

Figure \ref{fig: Results whole experiment} shows the results of the three day experiment. Note that the time axis is shifted by 11 hours and 45 minutes, to virtually let the experiment start at midnight.
Plot (a) depicts the real heating demand of the NEST building in dashed blue, the forecast conducted at midnight for Level 1 in orange, and the forecasts for Level 2, which are conducted every 15 minutes, in transparent grey. The initial forecast (orange) predicts the trend of the heating demand well, confirming the results of \cite{BunningFelixBollingerAndrewHeerPhilippSmithRoy2018}. The correction based on error-autocorrelation is visible, whenever the previous forecast significantly differs from the measured heating demand. This is the case at 00:00 at the beginning of the second day for example. At this point, the initial forecast of 23 kW (orange line) differs from the measured demand of 20 kW after one interval. The next corrected forecast at 00:15 (cyan line) therefore starts at 20 kW. The following corrected forecast at 00:30 (red line) starts at the measured demand of 16 kW. Both corrected forecasts merge back into the initial forecast over the course of the day because the error autocorrelation also decreases with time. It can also be seen that the experiment covers a range of heating demands from 10 kW, which is below the minimum thermal capacity of the heat pump, to 45 kW, which is close to the maximum capacity of the heat pump using only one compressor stage.

Plot (b) shows the reserves offered during the experiment. The reserves are either zero, or between 0.4 kW and 2.3 kW, which are the lower and upper limits. The upper limit is set as a constraint in the optimization problem, but is also a result of offering symmetric reserves and the electrical capacity of the heat pump being in the range of 8.2 to 12.8 kW. During the first day very little reserve is offered. This is for two reasons. First, the heat pump is frequently switched off because the demand is low. (See also plot (d) in Figure \ref{fig: Results whole experiment}). In this case constraint (\ref{eq:constraint4_2}) forces the corresponding elements in $D_w$ and $D_{\delta,e}$ to be zero, which means that recourse on uncertainties is no longer possible. Second, whenever the heat pump is on, it operates at the lower capacity limits, which for symmetric reserves results in offering no reserves. During the second and the third day of the experiment, reserves are offered during most of the 15 minute intervals. On day 1, 3.1\% of the electricity consumed is flexible, on day 2 and 3, 14.9\% and 19.1\% are flexible respectively. The average of all three days is 13.4\%. In addition to being constrained by the thermal inertia of the storage, the reserves are also limited by the region of operation of the heat pump, specifically to 2.3 kW in this particular case. On days one, two and three 4\%, 29\% and 41\% of this potential are exploited respectively. Considering the small size of the buffer storage compared to the thermal demand - on a day with an average heating demand of 35 kW, the thermal capacity of the buffer is just 3\% of the daily demand - this is a considerable outcome. As we will see in Section \ref{sec: Numerical case study and results}, it is mostly thanks to the use of affine policies. 

Plot (c) shows the average storage temperature in solid black, the temperature constraints at 28 \degree C and 38 \degree C in dotted black and the six temperature measurements at different heights of the storage tank in transparent colors. The average temperature stays between the constraints for most of the time, except one 30 minute instance between 7.30 and 8.30 on day two. However, during this time the heating demand of the building could still be served, as the upper temperature layer in the storage tank (transparent blue) was above 28 \degree C at all times, which is a result of the average storage temperature being a lower bound for the temperature of the top water layer in the storage tank. 
The results also show that, while this bound adds a layer of robustness, it is relatively tight at the constraints. This can be seen on day 2 at 14:00 and 23:00, for example, where the upper temperature layer approaches the lower constraint together with the average temperature. The reason is that the return temperature of the building is usually equal to the lower temperature constraint (and the water supplied by the heat pump has the same temperature as the upper temperature constraint of the tank).
Close investigation of the different temperature layers also shows that swapping of temperature layers indeed does not occur and the assumption for the storage model holds. The average temperature stays relatively close to the lower constraint most of the time as a result of the optimization: unless needed for reserves, temperatures above the minimum mean unnecessary consumption of electricity. During the first day, where the heating demand is below the minimum thermal capacity of the heat pump, this results in a hysteresis-like behaviour because the heat pump is switched on and off frequently. Between 8.00 and 12.00 on the second day, when the heat pump is switched on due to the reserves offered, the controller regulates the base load $u^0$ such that the temperature tracks the lower constraint. However, at times where the heating demand of the building is substantially overestimated by the initial demand forecast, the average storage temperature rises. This can be seen in the period between 12.00 and 18.00 on the third day.

Plot (d) of Figure \ref{fig: Results whole experiment} shows the set point for the electrical power of the heat pump in orange, the actual measured power of the heat pump in dashed blue and the possible range of the power due to the offered reserves in dotted red. As the results are difficult to read in this scale, an excerpt of this plot (9.00 to 11.00 of the second day) is shown in the subplot (a) of Figure \ref{fig: Results tracking performance}. It can be seen that the heat pump delivers the reserves offered by modulating the electricity between the limits defined by the base load and the offered reserves (dotted red) following the regulation signal depicted in Figure \ref{fig: Results tracking performance}(b). The effects of different sizes of offered reserves can also be observed. From 9.00 to 9.30, where no reserves are offered, the heat pump does not exactly follow the set point; the resulting tracking error is also evident in subplot (c) of Figure \ref{fig: Results tracking performance}. There are two reasons for this tracking error. First, there is measurement noise of approximately $\pm 200$ W. Second, the heat pump's internal controller only accepts integer set points for relative compressor speeds (e.g. 34\% and 35\%, but not 34.5\%), which leads to a discontinuous control signal. After 9.30, a range of different reserves is offered, visible from the span between the red dotted lines. Visible from 10.00 to 10.15, large reserves  lead to bigger tracking errors because of the ramping limits of the heat pump. Large tracking errors also occur when large steps in the base set-point for the heat pump appear (at 10.15).

Despite these tracking errors, the performance score\footnote{We note that the performance score is calculated with a Macro Excel sheet provided by PJM and could not be replicated with our own calculations. We refer the interested reader to the original source \cite{PJM}.} of the TSO PJM \cite{PJM} is better when higher amounts of reserves are offered. This is shown in subplot (d) of Figure \ref{fig: Results tracking performance}. Here, the green line depicts the \textit{delay score} (time delay between reserve signal and system response), the orange line depicts the \textit{accuracy score} (correlation between the reserve signal and the system response), the red line depicts the \textit{precision score} (error between reserve signal and system response) and the dashed blue line depicts the \textit{composite score} (average of the three). The scores are averaged over one hour intervals and are normalized in the interval $[0,1]$, with 1 being the best score. While the delay score is constantly high, both the accuracy score and especially the precision score become worse when the reserves offered are low, because the error relative to the offered reserves becomes large. Especially in the one-hour intervals where only small reserves are offered (or a combination of no reserves and small reserves), the composite performance score becomes low.

With respect to the whole experiment, this result is not problematic, as can be seen by going back to Figure \ref{fig: Results whole experiment}. In plot (e), the composite performance score is shown in blue. The orange line depicts the 20-hour moving average of the performance score. This is the metric used by PJM to judge whether a device or power plant is suitable for their reserve product. In the qualification phase, the limit for the performance score is 0.75 (dotted orange), while in the operational phase, the limit is lowered to 0.4 (dotted red). It can be seen that the performance score in this experiment is always well above both limits, confirming early results reported in \cite{Bunning2020a}.

\section{Numerical case study and results}
\label{sec: Numerical case study and results}

To better understand the performance of the control scheme, we have further analyzed it in two numerical experiments.

\subsection{Use of affine policies compared to open-loop MPC.}

In the first numerical experiment, we compare the solution of Level 1 with affine policies (as presented in Section \ref{sec: Methodology}) to a Level 1 scheme based on standard open-loop MPC without feedback policies for various constant heating demands $v$. The optimization problem for Level 1 for open-loop MPC is

\begin{subequations}
\label{probNoaffine}
\begin{alignat}{2}
&\!\min_{\substack{x,r,u^0,u_\text{th}, \\ z,\tilde{z},\epsilon}}         &\qquad& {f^\text{el}}^\top u^0 - {f^{r}}^\top r + \lambda ^\top \epsilon \label{eq:optProb2}\\
&\text{subject to} &      & x = A x_0 + B (u_\text{th}-v+\delta +e),\label{eq:constraint1_4}\\
&                  &      & u_\text{th}=  \alpha_\text{COP}(u^0+\bar{w}\odot r), \\
&				   & 	  & X_\text{min} - \epsilon \leq x \leq X_\text{max} + \epsilon \label{eq:constraint3_4}, \\
&				   &      & z U_\text{min} \leq u^0+w\odot r \leq z U_\text{max}, \\
&			    	   &      & \tilde{z}R_{min} \leq r \leq \tilde{z}R_{max}, \label{eq:constraint5_4}\\
&				   &      & z, \tilde{z} \in \mathbb{Z}^N_2 , \\
&				   &	  & \epsilon \geq 0, \\
&				   &	  & \forall w\in W, \forall \bar{w}\in \bar{W}, \forall \delta \in \Delta, \forall e \in E.
\end{alignat}
\end{subequations}

\begin{figure}[htb]
%	\capstart
	\centering
		\includegraphics[width=0.49\textwidth, trim=0 3mm 0 0,clip]{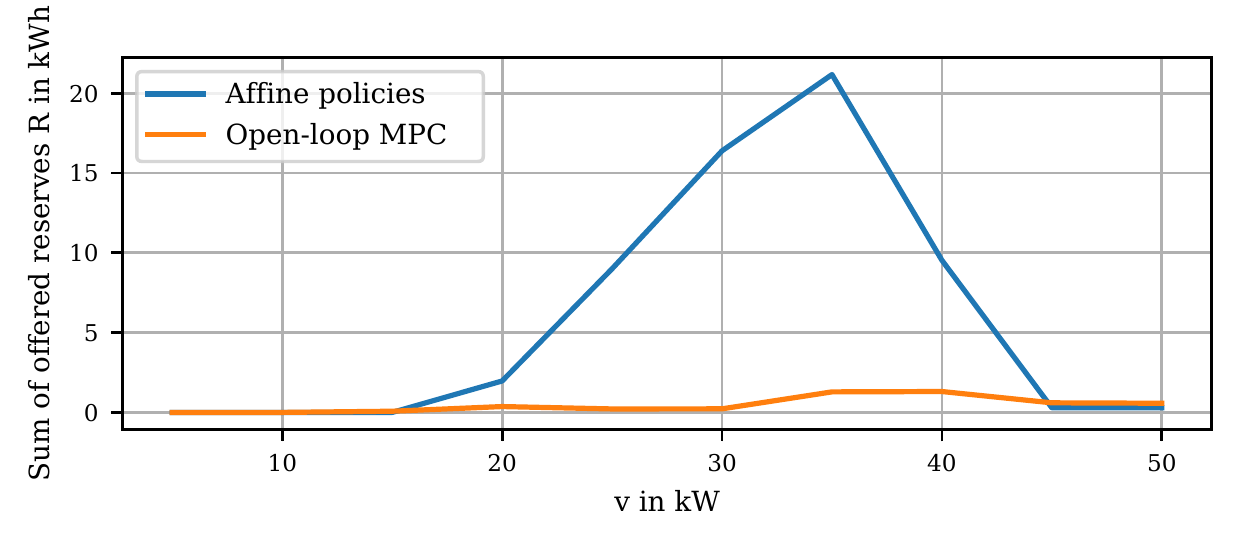}
	\caption{Reserves offered in Level 1: affine policies vs. open-loop MPC}
	\label{fig: Offered reserves in Level 1: affine policies vs. no affine policies}
\end{figure}

\begin{figure}[htb]
%	\capstart
	\centering
		\includegraphics[width=0.49\textwidth, trim=0 3mm 0 0,clip]{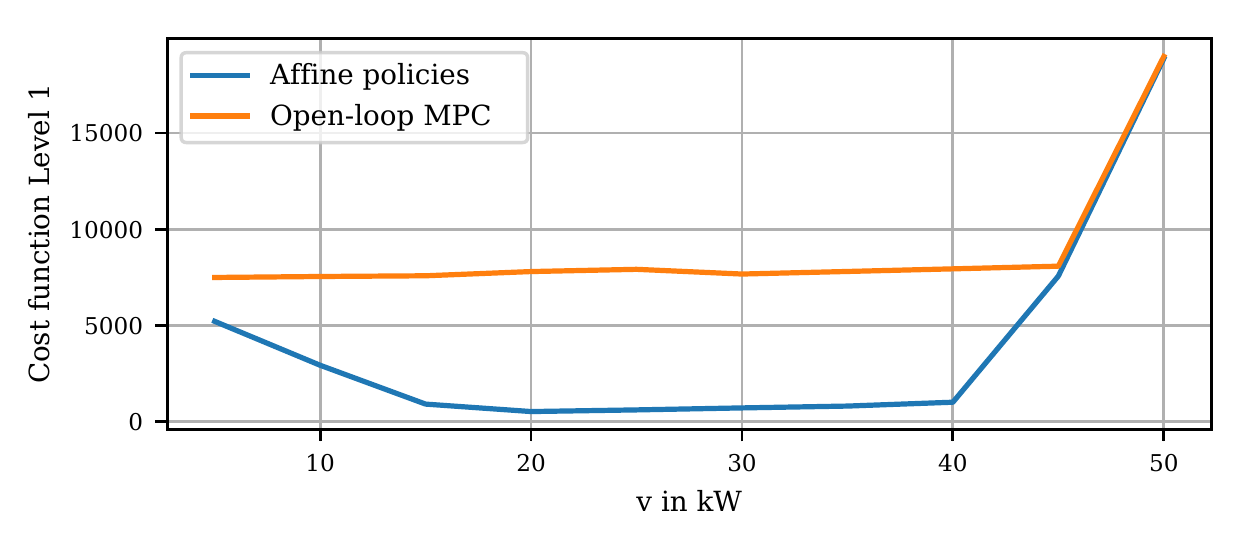}
	\caption{Cost in Level 1: affine policies vs. open-loop MPC}
	\label{fig: Cost in Level 1: affine policies vs. no affine policies}
\end{figure}

The comparison is made for constant heating demands between 5 kW and 50 kW, in steps of 5 kW. All parameters for the optimization schemes are the same as in Section \ref{sec: Experimental case study and results}. Figure \ref{fig: Offered reserves in Level 1: affine policies vs. no affine policies} and Figure \ref{fig: Cost in Level 1: affine policies vs. no affine policies} show the reserves offered and the cost functions (including the $\lambda$ term) for both approaches. It can be seen in Figure \ref{fig: Offered reserves in Level 1: affine policies vs. no affine policies} that without using feedback policies, the reserves offered are close to zero for most demands. This is due to the build-up of uncertainty in the state $x$ (storage temperature) over the horizon of the optimization problem in combination with a relatively small buffer, compared to the magnitude of the heating demand. In contrast, Level 1 with affine policies is able to offer reserves in most cases except when approaching the upper and lower capacity limits of the heat pump. From Figure \ref{fig: Cost in Level 1: affine policies vs. no affine policies} it can be seen that the value of the cost function is significantly lower when affine policies are used for all heating demands below 45 kW. Above 45 kW, there is no difference because the heat pump will always work at maximum capacity. We note however, that the played-out costs (with the MPC re-optimizing every 15 minutes) would have different results for both cases, which significantly depend on the uncertainty realizations. While the played-out behaviour would certainly be an interesting result to study, the computational effort is not feasible.

\subsection{Use of affine policies compared to a system with perfect knowledge.}

In a second numerical experiment, we compare the performance of Level 1 using affine policies with an omniscient Level 1 solution that has perfect knowledge of all uncertainty realizations at the time of optimization. In this case, the optimization problem becomes

\begin{samepage}
\begin{subequations}
\label{probExact}
\begin{alignat}{2}
&\!\min_{\substack{x,r,u^0,u_\text{th}, \\ z,\tilde{z},\epsilon}}         &\qquad& {f^\text{el}}^\top u^0 - {f^{r}}^\top r + \lambda ^\top \epsilon \label{eq:optProb2}\\
&\text{subject to} &      & x = A x_0 + B (u_\text{th}-v+ (\delta +e)),\label{eq:constraint1_4}\\
&                  &      & u_\text{th}=  \alpha_\text{COP}(u^0+\bar{w}\odot r), \\
&				   & 	  & X_\text{min} - \epsilon \leq x \leq X_\text{max} + \epsilon \label{eq:constraint3_4}, \\
&				   &      & z U_\text{min} \leq u^0+w_{min}\odot r, \\
&				   &      & u^0+w_{max}\odot r \leq z U_\text{max}, \\
&			    	   &      & \tilde{z}R_{min} \leq r \leq \tilde{z}R_{max}, \label{eq:constraint5_4}\\
&				   &      & z, \tilde{z} \in \mathbb{Z}^N_2 , \\
&				   &	  & \epsilon \geq 0,
\end{alignat}
\end{subequations}
\end{samepage}

\noindent where $(\delta+e)$ is drawn from a uniform distribution with the same limits as the uncertainty set $E \oplus \Delta$, and $w_{min}$, $w_{max}$ and $\bar{w}$ are extracted from the regulation signal used (PJM RegD of 27th of January 2019). Here, $w_{min}$ and $w_{max}$ are the minimum and maximum value of the regulation signal that occurs during a 15 minute interval respectively; $\bar{w}$ is the average of the interval.

The experiment is conducted for a constant heating demand $v$ between 5 kW and 50 kW, in steps of 5 kW, with 10000 uncertainty realizations for each $v$. Optimization problem (\ref{probExact}) is solved much faster than the robust counterpart of problem (\ref{prob2}), requiring less than 5 seconds for convergence.

\begin{figure}[htb]
%	\capstart
	\centering
		\includegraphics[width=0.49\textwidth, trim=0 3mm 0 0,clip]{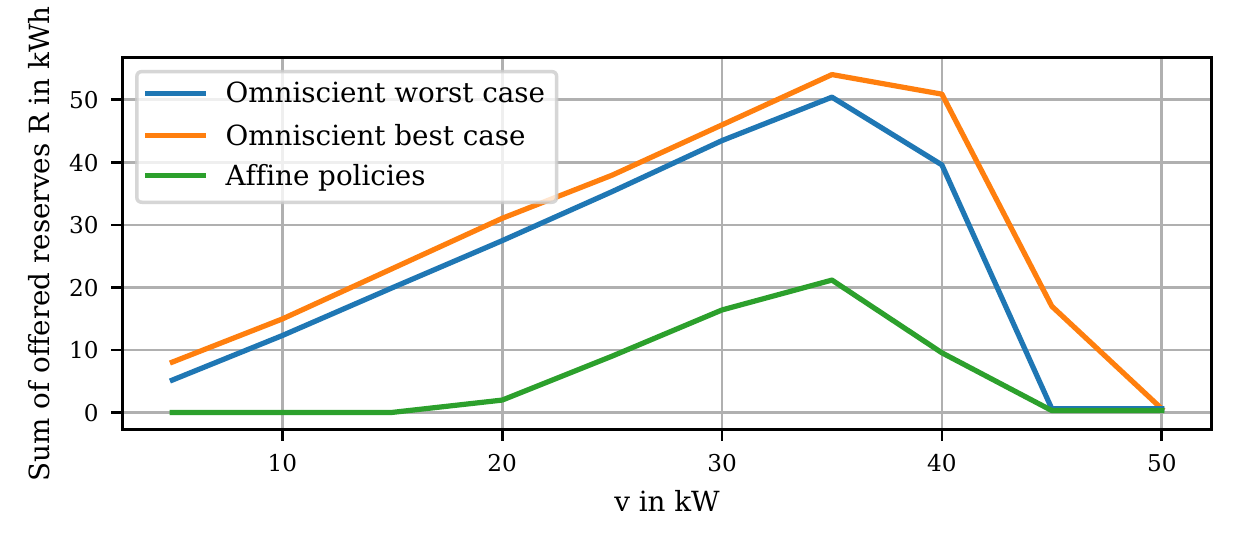}
	\caption{Reserves offered in Level 1: affine policies vs. perfect knowledge}
	\label{fig: Offered reserves in Level 1: affine policies vs. perfect knowledge}
\end{figure}

\begin{figure}[htb]
%	\capstart
	\centering
		\includegraphics[width=0.49\textwidth, trim=0 3mm 0 0,clip]{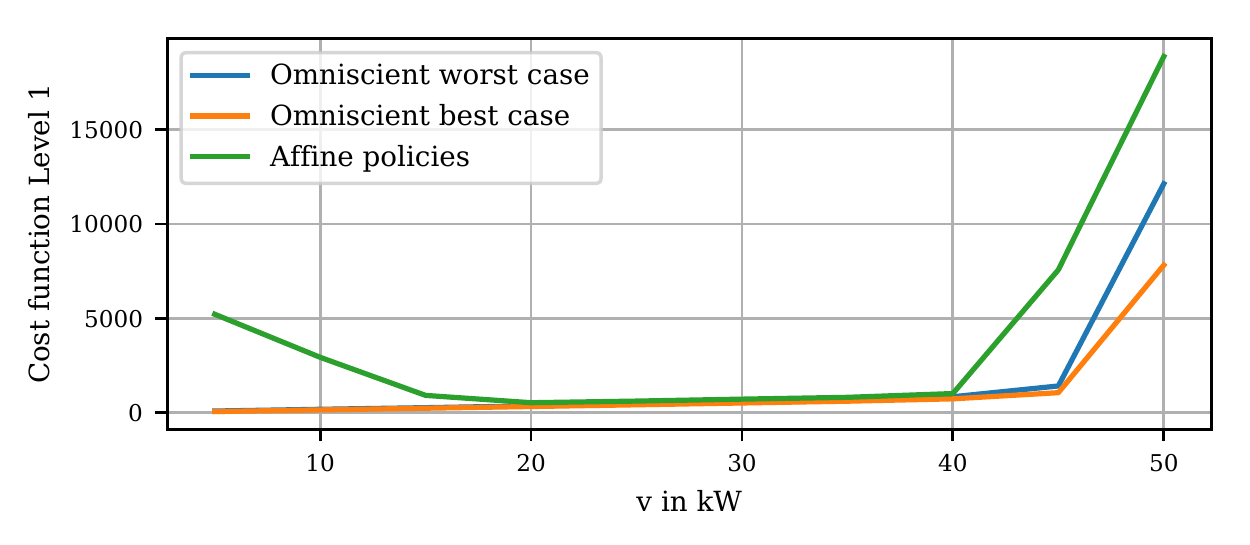}
	\caption{Cost in Level 1: affine policies vs. perfect knowledge}
	\label{fig: Cost in Level 1: affine policies vs. perfect knowledge}
\end{figure}

\begin{figure}[htb]
%	\capstart
	\centering
		\includegraphics[width=0.49\textwidth, trim=0 3mm 0 0,clip]{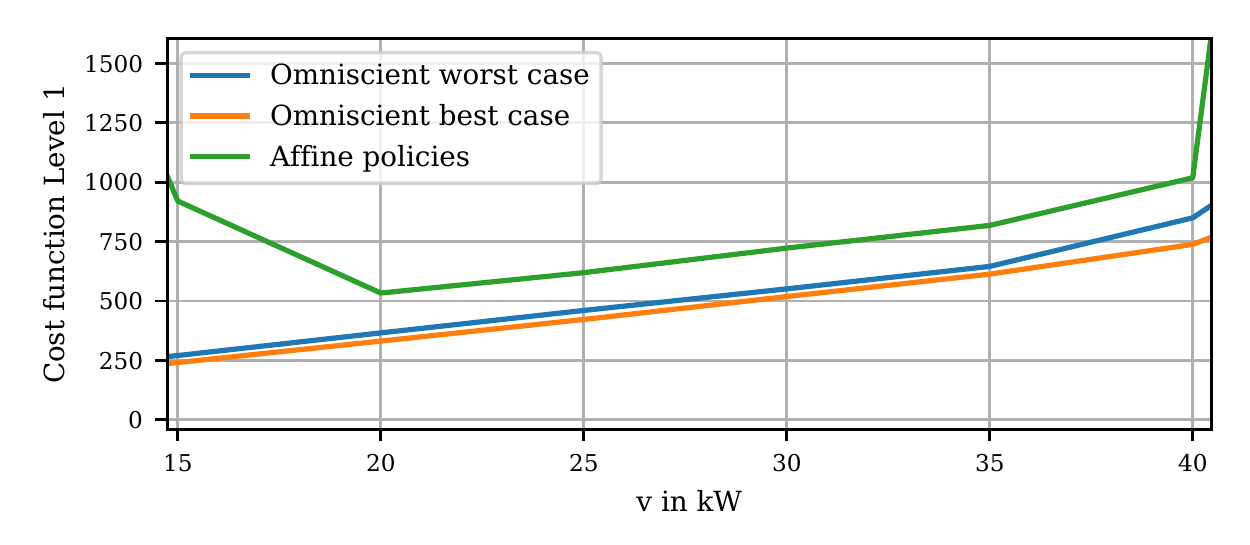}
	\caption{Cost in Level 1: affine policies vs. perfect knowledge, in detail}
	\label{fig: Cost in Level 1: affine policies vs. perfect knowledge, in detail}
\end{figure}

Figure \ref{fig: Offered reserves in Level 1: affine policies vs. perfect knowledge} shows the reserves offered for both cases for all different $v$. Here, the orange line denotes the best result achieved with the omiscient system with 10000 uncertainty realizations, while the blue line denotes the worst result. Each set of uncertainty realizations drawn leads to a different cost function and to differing amounts of reserves offered, as each realization either benefits the operation or harms it, compared to the case without uncertainty. We report the best case and worst case among these realizations. The green line shows the solution of Level 1 with affine policies. It can be seen that for all cases except $v=45$ and $50$ kW, the solutions with perfect knowledge of the uncertainties, offer significantly more reserves than the scheme with affine policies. The results for the cost function, depicted in Figure \ref{fig: Cost in Level 1: affine policies vs. perfect knowledge} for all $v$ and depicted in Figure \ref{fig: Cost in Level 1: affine policies vs. perfect knowledge, in detail} for $v$ between 15 and 40 kW, show that the scheme with perfect knowledge performs significantly better at very high and very low heating demands, which is due to avoiding the use of the slack variable $\epsilon$. Also at intermediate demands, there is an offset between the cost functions of both schemes. This can be explained by the fact that the scheme with perfect knowledge is able to operate right at the storage temperature constraints (maximizing offered reserves or minimizing the base load), while the scheme with affine policies needs to stay at least $B(\Delta \oplus E)$ away from the storage temperature constraint, even when no reserves are offered. This is the case because affine policies can only react to uncertainties in the timestep after their realization, unlike a system with perfect knowledge, which can plan ahead and can compensate for uncertainties before they are revealed. The larger optimality gap at 15 kW can again be explained by the fact that affine policies can only work when $z$ is non-zero, which is often not the case for low heating demands.

\section{Limitations and future directions}
\label{sec: Limitations}

The numerical results show that using affine policies on uncertainties significantly increases reserve provision and decreases cost when compared to using open-loop MPC. Nevertheless, the comparison with the omniscient optimization indicates that there is still room for improvement. One possibility could be the use of more sophisticated policies (for example deflected linear \cite{Chen}, or piecewise linear \cite{Georghiou2015} policies). However, due to causality constraints no policy can match the performance of an omniscient controller that knows the future. 

Moreover, in future work the presented method of only exploiting the thermal inertia of buffer storage should be extended to and compared to a scheme where the inertia of a subset of the connected buildings (willing to participate in the reserve provision scheme) is also taken into account, to further investigate the added potential by including building thermal dynamics in the problem formulation. Configurations where the heat pump is supported by additional heat sources, such as Combined Heat and Power units, should also be studied as their flexibility could potentially increase the reserves provision.  

On a more practical note, a limitation of the approach presented here is that it can only work with variable speed heat pumps, which, although becoming more common, are still relatively rare compared to fixed speed heat pumps. To offer reserves with fixed speed devices, an aggregation mechanism that pools heat pumps of many different buildings would still be necessary in order to follow a continuous reserve signal. A further limitation for air-sourced heat pumps could be the requirement of regular de-freezing cycles. However, as these do not last long and our achieved performance scores are well above the qualification limit, they could be integrated without a change to the control scheme.

\section{Conclusion}
\label{sec: Conclusion}

In this work, we have combined a three-level control scheme based on robust optimization with affine policies with heating demand forecasting based on ANN and online correction methods with the aim of offering frequency regulation reserves with heat pumps and water storage in buildings and district heating settings. The approach works without the necessity of a dynamic building model, which reduces modeling effort compared to including reserve provision in MPC building temperature control. Moreover, it alleviates privacy concerns and has reduced hardware requirements, as no indoor temperature or occupancy measurements are necessary to guarantee indoor comfort. 

The experiments on the heat pump and water storage in the NEST demonstrator building have shown that the three-level control approach with affine policies on uncertain variables presented here is viable. The method allows the offering of frequency regulation reserves with a single variable speed heat pump and a small (compared to the overall heating demand) buffer storage. On average, 13.4\% of the consumed electricity is flexible as a result of the reserves offered. The performance of tracking the regulation signal is more than sufficient. Our experiments indicate that large buildings or district heating systems equipped with variable speed heat pumps can in principle directly be used for ancillary services, without the need for aggregation, and that the control scheme is robust in practical applications. The heating demand forecasting approach with ANN and online correction methods gives predictions with high accuracy, such that the demand of the building can always be met. The numerical case studies show that, while there is still a performance gap compared to an omniscient controller, the use of affine policies significantly reduces costs and improves reserve provisions, compared to open-loop MPC. In future work, the method should be extended and compared to a problem formulation that also exploits the thermal inertia of connected buildings.

% use section* for acknowledgment
\section*{Acknowledgment}

We would like to thank Kristina Orehounig and Viktor Dorer for their valuable help and support. We are also grateful to Ahmed Aboudonia, Annika Eichler, Benjamin Flamm, Reto Fricker, Marc Hohmann, Benjamin Huber, Mathias Hudoba de Badyn, Andrea Ianelli, Mohammad Khosravi, Ralf Knechtle, Francesco Micheli, Anil Parsi and Sascha Stoller for fruitful discussions. This research project is financially supported by the Swiss Innovation Agency Innosuisse and is part of the Swiss Competence Center for Energy Research SCCER FEEB{\&}D.

\section*{Conflict of interest}

None declared.

\printcredits

%% Loading bibliography style file
%\bibliographystyle{model1-num-names}
\bibliographystyle{cas-model2-names}

% Loading bibliography database
\bibliography{cas-refs}

%\vskip3pt
%
%\bio{}
%Author biography without author photo.
%Author biography. Author biography. Author biography.
%Author biography. Author biography. Author biography.
%Author biography. Author biography. Author biography.
%Author biography. Author biography. Author biography.
%Author biography. Author biography. Author biography.
%Author biography. Author biography. Author biography.
%Author biography. Author biography. Author biography.
%Author biography. Author biography. Author biography.
%Author biography. Author biography. Author biography.
%\endbio
%
%\bio{figs/pic1}
%Author biography with author photo.
%Author biography. Author biography. Author biography.
%Author biography. Author biography. Author biography.
%Author biography. Author biography. Author biography.
%Author biography. Author biography. Author biography.
%Author biography. Author biography. Author biography.
%Author biography. Author biography. Author biography.
%Author biography. Author biography. Author biography.
%Author biography. Author biography. Author biography.
%Author biography. Author biography. Author biography.
%\endbio
%
%\bio{figs/pic1}
%Author biography with author photo.
%Author biography. Author biography. Author biography.
%Author biography. Author biography. Author biography.
%Author biography. Author biography. Author biography.
%Author biography. Author biography. Author biography.
%\endbio

\end{document}